\documentclass[12pt,a4]{article}
\usepackage{graphicx}
\usepackage[numbers]{natbib}
\usepackage{pslatex}
\usepackage{color}
\usepackage{algorithm}
\usepackage{algorithmic}
\usepackage{amsmath,marvosym,enumerate,multirow,amsfonts}
\usepackage{verbatim}
\usepackage{fancyvrb}
\usepackage{fancybox}
\usepackage{setspace}

\providecommand{\keywords}[1]{\textbf{\textit{Keywords:}} #1}
\newtheorem{example}{Example}
\setcitestyle{super}

\definecolor{lightgray}{gray}{0.95}

\newenvironment{MyVerbatim}{\VerbatimEnvironment%
    \noindent\begin{Sbox}\footnotesize
        \begin{minipage}{\dimexpr\linewidth-20\fboxsep-40\fboxrule}
          \begin{Verbatim}
}{%
      \end{Verbatim}%
      \end{minipage}%
      \end{Sbox}%
      \fcolorbox{black}{lightgray}{\TheSbox}%
}

\begin{document}
\title{Numerical Simulation of Multi-phase Flow in Porous Media on Parallel Computers}

\author{Hui Liu\thanks{Authors to whom Correspondence may be addressed. Email addresses: hui.jw.liu@gmail.com}, Lihua Shen, Kun Wang, Bo Yang, Zhangxin Chen\\
Department of Chemical and Petroleum Engineering \\
University of Calgary, 2500 University Drive NW, Calgary, AB, Canada
}

\date{}
\maketitle

{\noindent \rule{\linewidth}{0.5mm}}

\bigskip

\section*{Abstract}
A parallel reservoir simulator has been developed, which is designed for large-scale black oil simulations. It handles three phases, including water, oil and gas, and three components, including water, oil and gas. This simulator can calculate traditional reservoir models and naturally fractured models. Various well operations are supported, such as water flooding, gas flooding and polymer flooding. The operation constraints can be fixed bottom-hole pressure, a fixed fluid rate, and combinations of them. The simulator is based on our in-house platform, which provides grids, cell-centred data, linear solvers, preconditioners and well modeling.  The simulator and the platform use MPI for communications among computation nodes.  Our simulator is capable of simulating giant reservoir models with hundreds of millions of grid cells.  Numerical simulations show that our simulator matches with commercial simulators and it has excellent scalability.

\bigskip
\noindent \keywords{Reservoir Simulation, Black oil, Oil-water, Parallel Computing}

\section{INTRODUCTION}
Reservoir simulations are powerful tools for petroleum engineers. Simulators can be used
to model oil, gas and water flow underground and interactions between reservoirs and wells,
which can predict well performance, such as oil rates, gas rates, water rates and bottom hole pressure.
These tools are employed to validate and optimize well operations.
Inputs of reservoir simulators include properties of fluids and reservoir rock.
Geological models of reservoirs can be obtained from seismic imaging. Usually they are complex and
highly heterogeneous, which introduce numerical difficulties for reservoir simulators.
When a reservoir model is large enough, a typical simulator may take days or even longer
to complete one simulation study case, especially in thermal simulations.
Effective numerical methods,
linear solvers, and fast computing techniques need be studied.

Reservoir simulation has been a popular research topic for decades.
Various reservoir models and well treatments as well as their numerical methods \citep{Book-Chen} and
fast computer techniques have been developed \citep{phg,plat,pennsim}.
Chen et al. studied finite element methods and finite difference methods
for the black oil, compositional and thermal models \citep{Book-Chen}. Newton methods,
linear solvers and preconditioners were also studied \citep{Book-Chen}.
Kaarstad et al. \citep{PS-Kaa} implemented a parallel two-phase oil-water simulator.
Rutledge et al. \citep{PS-Rut} implemented a compositional simulator,
which was designed for parallel computers. The simulator used the IMPES (implicit pressure-explicit saturation) method
and only pressure was solved for.
Shiralkar and his collaborators \citep{PS-Shi} developed a portable parallel reservoir simulator, which
could run on a variety of parallel systems.
Killough et al. \citep{PS-Kil2} studied locally refinement techniques,
which could improve accuracy and reduce calculations compared with global grid refinement.
The technique is useful to complex reservoir models, such as in-situ combustion.
Dogru and his group \citep{PS-Dogru2} developed parallel simulators using
structured and unstructured grids, which could handle faults, pinchouts, complex wells, polymer
flooding in non-fractured and fractured reservoirs.
Their parallel simulator was highly efficient and scalable, and
it could compute giant reservoir models with billions of grid cells.
Zhang et al. developed a general-purpose parallel platform for large-scale scientific applications.
The platform was designed for adaptive finite element and adaptive finite
volume methods \citep{phg,phg-quad}, and it used tetrahedral grids. The newest-vertex based bisection refinement
method, various linear solvers, preconditioners and eigenvalue solvers were provided.
The package has been applied to black oil simulations using discontinuous Galerkin methods \citep{kwang}.
Chen and his group developed a parallel platform to support the study of large-scale reservoir simulations
and this platform was implemented to support black oil, compositional, thermal,
and polymer flooding models \citep{plat, jcp-bos, chenpoly}.
Guan and his collaborators implemented a parallel simulator, which could compute black oil model
and compositional model \citep{pennsim}. Massive reservoir models with hundreds of millions
of grid cells were reported using 10,000 MPIs \citep{pennsim}.
Wheeler has developed a parallel black oil simulator \citep{mary} and she studied numerical methods, linear solvers,
and preconditioner techniques.
It is well-known that solution of linear systems
from black oil simulations occupies most of running time.
Many preconditioning methods have been proposed and applied to reservoir simulations,
such as constrained pressure residual (CPR) methods \citep{CPR-old,CPR-cao},
multi-stage methods \citep{Study-Two-Stage}, multiple level preconditioners
\citep{mlp}, fast auxiliary space preconditioners (FASP) \citep{FASP}, and a family of parallel CPR-like
methods \citep{bos-pc}.

A black oil simulator has been implemented, which handles several different models, such as the
standard three-phase black oil model, two-phase oil-water model, and
dual-porosity and dual permeability model in naturally fractured reservoirs.
The implementation details are introduced in this paper, including models, numerical methods,
and parallel implementations.
Numerical experiments show that our simulator can match commercial simulators and
it has excellent scalability.
Our simulator can compute large-scale reservoir simulation problems with
hundreds of millions of grid cells.

\section{RESERVOIR SIMULATION MODELS}
\label{model}

We assume the fluid in black oil model satisfies Darcy's law, which establishes a relationship among flow rate,
reservoir properties, fluid properties and phase pressure differences, and it is written as:
\begin{equation}
Q = - \frac{\kappa A\Delta p}{\mu L},
\end{equation}
where, $\kappa$ is permeability of rock (or reservoir),
$A$ is cross-section area in some direction,  $\Delta p$ is pressure difference, $\mu$
is viscosity of the fluid, and $L$ is length of a porous media in the direction.
Its differential form is written as:
\begin{equation}
q = \frac{Q}{A} = - \frac{\kappa}{\mu} \nabla p.
\end{equation}

\subsection{Black Oil Model}

The black oil model is isothermal, and it has three components (water, oil and gas) and
three phases (water, oil and gas). The gas component can exist in oil phase (solution gas)
and gas phase (free gas). The water component exists only in water phase and the oil component
exists in oil phase only.

By applying Darcy's law, three mass conservation equations for three components in non-fractured reservoirs
are written as:
\begin{equation}
 \left\{
 \begin{aligned}
& \frac{\partial}{\partial t} (\phi s_o \rho_o^o)& = &\nabla \cdot (\frac{K K_{ro}}{\mu_o} \rho_o^o \nabla \Phi_o) + q_o, \\
& \frac{\partial}{\partial t} (\phi s_w \rho_w)& = &\nabla \cdot (\frac{K K_{rw}}{\mu_w} \rho_w \nabla \Phi_w) + q_w, \\
& \frac{\partial(\phi \rho_o^g s_o + \phi \rho_g s_g)}{\partial t}& = &
\nabla \cdot ( \frac{K K_{ro}}{\mu_o} \rho_o^g \nabla \Phi_o) + q_o^g \\
& &  + & \nabla \cdot ( \frac{K K_{rg}}{\mu_g} \rho_g \nabla \Phi_g)  + q_g,
\end{aligned}
 \right.
\end{equation}
where, any phase $\alpha$ $(\alpha = o,w,g)$,
    $\Phi_{\alpha}$ is its potential, $\phi$ and $K$ are porosity and permeability, and
      $s_{\alpha}$, $\mu_{\alpha}$, $p_{\alpha}$,
    $\rho_{\alpha}$, $K_{r\alpha}$ and $q_{\alpha}$ are
    phase saturation, phase viscosity, phase pressure,
    phase density, relative permeability and well rate, respectively.
$\rho_o^o$ and $\rho_o^g$ are density of oil component in oil phase
and density of solution gas in oil phase.     They properties have the following constraints:
\begin{equation}
\left\{
\begin{aligned}
& \Phi_{\alpha} = p_{\alpha} + \rho_{\alpha} G z, \nonumber \\
& s_o + s_w + s_g = 1, \nonumber\\
& p_w = p_o - p_{cow} (s_w),\nonumber \\
& p_g = p_o + p_{cog} (s_g) ,\nonumber
\end{aligned}
\right.
\end{equation}
where $G$ is the gravitational constant and $z$ is the reservoir depth.
In this paper, a no-flow condition is applied as the boundary condition.
The relative permeabilities of water phase, gas phase and oil phase, $K_{rw}$, $K_{ro}$ and $K_{rg}$
are functions of water saturation and gas saturation,
\[
\left\{
\begin{aligned}
& K_{rw} = K_{rw}(S_w), \\
& K_{rg} = K_{rg}(S_g), \\
& K_{ro} = K_{ro}(S_w, S_g),
\end{aligned}
\right.
\]
where $K_{ro}$ is calculated by using the Stone II formula. In our simulator,
$K_{rw}$ and $K_{ro}$ are input parameters.
The capillary pressures are also input parameters, which are functions of water and gas saturations.
Usually oil phase pressure, water saturation, gas saturation (or bubble point pressure)
are chosen as unknowns.

For fractured reservoirs, each grid cell is divided into matrices and fracture, and each matrix and fracture have its
own pressure, saturation, and conservation laws. The commonly used models are dual porosity model, dual permeability
model and MINC (multiple interacting continuum) model.
The mass conservation laws are similar to non-fractured reservoirs except that transfer
terms should be defined among matrices and fractures.

\subsection{Two-Phase Flow Model}

The two-phase model ignores gas phase and can be read as a simplified model of the standard black oil model,
 \citep{Book-Chen}:
\begin{equation}
\label{eq-two}
 \left\{
 \begin{aligned}
& \frac{\partial}{\partial t} (\phi s_o \rho_o) & = & \nabla \cdot ( \frac{K K_{ro}}{\mu_o} \rho_o \nabla \Phi_o) + q_o \\
& \frac{\partial}{\partial t} (\phi s_w \rho_w) & = & \nabla \cdot ( \frac{K K_{rw}}{\mu_w} \rho_w \nabla \Phi_w) + q_w.
\end{aligned}
 \right.
\end{equation}

The variables are the same as black oil model and they are the following constraints,
\begin{equation}
 \left\{
 \begin{aligned}
&\Phi_{\alpha} = p_{\alpha} + \rho_{\alpha} G z, \nonumber \\
& s_o + s_w  = 1, \nonumber\\
& p_w = p_o - p_{cow} (s_w),\nonumber
\end{aligned}
\right.
\end{equation}
and again, $G$ is the gravitational constant and $z$ is the reservoir depth. In our simulator,
oil phase pressure and water saturation are chosen as unknowns.

\subsection{Well Management}

The source-sink model and Peaceman method \citep{PWM} are adopted to manage well operations.
For each perforation block $m$,
its well rate, $q_{\alpha,m}$, is calculated as:
\begin{equation}
q_{\alpha,m} = W_i \frac{\rho_\alpha K_{r\alpha}}{\mu_\alpha} (p_{h} - p_\alpha - \rho_\alpha \wp (z_{h} - z)),
\end{equation}
where $p_{h}$ is bottom hole pressure of a well, $W_i$ is well index of the perforated block $m$,
$z_{h}$ is reference depth for bottom hole pressure $p_{h}$, $z$ is depth of the perforated block $m$,
and $p_\alpha$ is phase pressure of interested phase. Bottom hole pressure should be known when calculating
well rate.

Various well operation strategies may be applied to a well at different time stage,
such as fixed bottom hole pressure operation, fixed oil rate operation, fixed water
rate operation or fixed liquid rate operation.

When the fixed bottom hole pressure operation is applied to a well,
$p_{h}$ is known and keeps unchanged. Since the phase pressure is known,
the well rate $q_{\alpha, m}$ is known.
The constraint equation for the well is
\begin{equation}
p_h = c,
\end{equation}
where $c$ is a constant set by the user input. In this case, there is no unknown to be solved
for the well.

When a fixed rate operation is applied to a well, its
bottom hole pressure is an unknown, and a mass conservation equation
for the well should be included. For the fixed water rate operation,
the constraint is
\begin{equation}
\sum_{m} {q_{w,m}} = q_w,
\end{equation}
where $q_w$ is constant.
For the fixed oil rate operation, its mass conservation equation is
\begin{equation}
\sum_{m} {q_{o,m}} = q_o,
\end{equation}
where $q_o$ is constant. For the fixed liquid rate condition, the constraint equation is
\begin{equation}
\sum_{m} ({q_{o,m} + q_{w,m}}) = q_o + q_w.
\end{equation}

Different constraints and combinations of them may be applied to a well at different time stages,
so a scheduler should be included in a simulator, which can detect operation changes.

\section{NUMERICAL METHODS}
We focus on structured grids and finite difference method is applied to these models. Reservoir models
are highly coupled nonlinear systems. Newton method and inexact Newton method are employed to
solve the nonlinear systems. The standard Newton method solves linear system accurately while
inexact Newton method solves linear system approximately. In real-world simulations,
inexact Newton method can accelerate simulation and reduce running time.

\subsection{Nonlinear Methods}
The algorithm for inexact Newton method \citep{golub,inexact-Newton} is shown by Algorithm \ref{inewton-alg}.

\begin{algorithm}
\caption{The inexact Newton Method}
\label{inewton-alg}
\begin{algorithmic}[1]
\STATE Given an initial guess $x^0$ and stopping criterion $\epsilon$, let $l = 0$
 and assemble the right-hand side $b(x^l)$.
\WHILE{$\left\|b(x^l) \right\| \ge \epsilon$}
\STATE Assemble the Jacobian matrix $A$.
\STATE Find $\theta_l$ and $x$ such that
    \begin{equation}
    \label{inexact-newton-alg}
    \left\| b(x^l) - A \delta x \right\| \leq \theta_l \left\| b(x^l) \right\|,
    \end{equation}
\STATE    Let $l = l+1$ and $x^l = x^{l-1} + \delta x$.
\ENDWHILE
\STATE $x^* = x^l$ is the solution of the nonlinear system.
\end{algorithmic}
\end{algorithm}

The algorithm is the same as Newton methods except the choice of $\theta_l$.
The standard  Newton method chooses a small tolerance, such as 1e-7; in this case, the
solution of the corresponding linear system is accurate. The price is that
the linear solver occupies large part of overall simulation time.
The termination criteria of the inexact Newton method is larger compared with
standard Newton method, such as 1e-2. And also, its value is set automatically.
Three commonly-used choices are listed as follows \citep{inexact-Newton}:
\begin{equation}
\theta_l =
\left\{
\begin{aligned}
\label{choice-inm}
& \frac{\left\| b(x^l) - r^{l-1} \right\|}{\left\| b(x^{l-1}) \right\|}, \\
& \frac{\left\| b(x^l) \right\| - \left\| r^{l-1} \right\|}{\left\| b(x^{l-1}) \right\|}, \\
& \gamma \left( \frac{\left\| b(x^l) \right\|}{\left\| b(x^{l-1}) \right\|} \right)^{\beta},
\end{aligned}
\right.
\end{equation}
where $r^l$ is the residual of the $l$-th iteration,
\begin{equation}
r^l = b(x^l) - J \delta x.
\end{equation}

\subsection{Preconditioner}
A linear system, $Ax = b$, is derived from each Newton iteration, which
is un-symmetric and ill-conditioned. Krylov solvers are employed usually.
If a proper ordering technique is applied, the matrix $A$ has the following structure,

\begin{equation}
\label{mat-ab}
A = \left(
        \begin{array}{lll}
        A_{pp} \hspace{0.1cm}   & A_{ps} \hspace{0.1cm} & A_{pw}   \\
        A_{sp} \hspace{0.1cm}   & A_{ss} \hspace{0.1cm} & A_{sw}   \\
        A_{wp} \hspace{0.1cm}   & A_{ws} \hspace{0.1cm} & A_{ww}   \\
        \end{array}
        \right),
\end{equation}
where $A_{pp}$ is the {matrix coefficients} corresponding to the pressure unknowns,
$A_{ss}$ is the {matrix coefficients} corresponding to other unknowns, such as
saturations, bubble-point pressure, and polymer concentration, and $A_{ww}$ is the {matrix coefficients}
corresponding to {well bottom hole pressure}, and other matrices are coupled ones.

A decoupling operator, which is defined as a matrix $D$, is applied to $Ax = b$ and it
converts the original linear system to an equivalent one:
\begin{equation}
D^{-1} A x = D^{-1} b.
\end{equation}
The decoupled system is solved instead of original system.
Many decoupling strategies have been proposed, such as Quasi-IMPES method \citep{Quasi-IMPES} and ABF method \citep{ABF}.
The operators are simple and cheap to create. Here we introduce the Quasi-IMPES decoupling operator,
which is defined as
\begin{equation}
D_{QI} = \left(
        \begin{array}{ccc}
        I \hspace{0.1cm}   & D_{ps} D_{ss}^{-1} \hspace{0.1cm} & 0   \\
        0 \hspace{0.1cm}   & I \hspace{0.1cm} & 0   \\
        0 \hspace{0.1cm}   & 0 \hspace{0.1cm} & I   \\
        \end{array}
        \right),
\end{equation}
where $D_{ps} = diag(A_{ps})$ and $D_{ss} = diag(A_{ss})$.

The system is difficult to solve and many CPR-type preconditioners have been developed, such as classical CPR method and
FASP method.
We also designed a set of parallel CPR-type methods for black oil model and compositional model \citep{bos-pc}.
For the sake of completeness, one of them, CPR-FPF, is introduced, where F means to apply RAS method (Restricted
Additive Schwarz) \citep{RAS} to linear system $A r = f$, and P means to apply AMG method
 to solve linear $A_{pp} r_p = f_p$. RAS method and AMG method are well-known to be
 scalable for parallel computing. The algorithm CPR-FPF method
is described by Algorithm \ref{pc-fpf}.

\begin{algorithm}[!htb]
\caption{The CPR-FPF Preconditioner for preconditioning system $A x = f$.}
\label{pc-fpf}
\begin{algorithmic}[1]
\STATE $x = R(A)^{-1} f $.
\STATE {$r = f - Ax$}
\STATE $x = x + \varPi_p AMG(A_{pp})^{-1} \varPi_r  r $.
\STATE {$r = f - Ax$}
\STATE $x = x + R(A)^{-1} r $.
\end{algorithmic}
\end{algorithm}

The RAS method is one of domain decomposition methods, which is popular for parallel computing. Each process
setups a local problem, whose size is determined by local graph and overlap. Each local problem is solved
by a serial solver, such as ILU(k), ILUT(p, tol) and other methods. In our simulator, the default local
solver is ILU(0). There is no communication in the solution of local problem. The default overlap is one. The
algebraic multigrid solver we use is BoomerAMG from HYPRE \citep{HYPRE2}.

\subsection{Data structures and Algorithms}
An in-house parallel platform has been developed to support the implementations of parallel reservoir simulators.
The platform provides structured grid, cell-centered data, mapping, linear solvers, preconditioners, well modeling,
parallel input and output, keywords parsing and visualization \citep{plat}.

\begin{figure}[!htb]
    \centering
    \vspace{1.5cm}
    \includegraphics[width=0.5\linewidth]{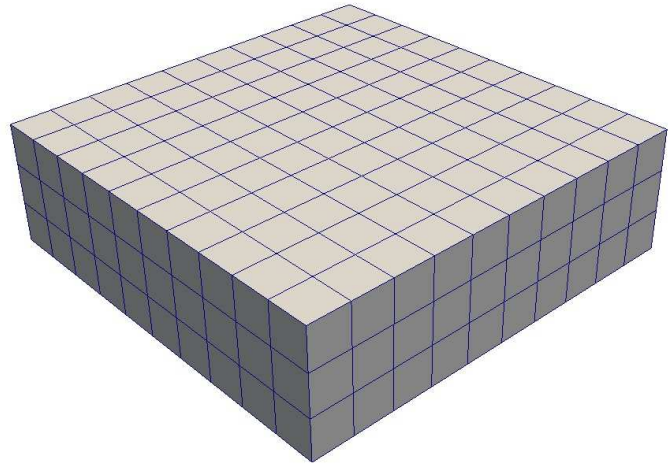}
    \caption{A structured grid}
    \label{fig-hex}
\end{figure}

Currently, regular structured hexahedral grids are provided, which have simple geometry and topology.
A structured grid is shown in Fig. \ref{fig-hex}.
Each cell of a grid is a hexahedron. Each cell has an
integer coordinate $(i, j, k)$ and each component (i, j, and k) is numbered
along x-, y- and z-axis. Each cell also has a unique global index and when they are distributed
in MPIs, each of them has a local index. Fig. \ref{fig-ds-cell} shows data structure
of \verb|CELL|, which stores geometric info, such as centroid coordinate (\verb|ctrd|),
face areas (\verb|area|), volume (\verb|vol|), index, integer coordinates in each direction and boundary type
of each face.

\begin{figure}[!htb]
\centering
\begin{MyVerbatim}
typedef struct CELL_
{
    COORD      ctrd;
    FLOAT      area[6];
    FLOAT      vol;

    void       *nb[6];
    INT        vert[8];
    INT        index;
    INT        idx[3];

    USHORT     bdry_type[6];

} CELL;
\end{MyVerbatim}
\caption{Data structure of CELL}
\label{fig-ds-cell}
\end{figure}

A grid is distributed in $N_p$ MPI tasks and each MPI task
owns a sub-grid. Let $\mathbb{G}$ be the grid, which has $N_g = n_x \times n_y \times n_z$ cells,
\begin{equation}
\mathbb{G} = \{C_1, C_2, \cdots, C_{N_g}\},
\end{equation}
where $C_i$ is the $i$-th cell of $\mathbb{G}$.
Let $\mathbb{G}_i$ be the sub-grid owned by the $i$-th MPI task. For any cell,
its neighboring cells may belong to different sub-grids. When we discretize
black oil models, information from neighboring cell is required, then
communication pattern can be modeled by dual graph and communication volume can
be approximated by cutting edges. In our simulator, the Hilbert space-filling curve method is employed
to partition a grid.

Cell-centered data module is designed to support finite difference methods and
finite volume methods. The platform also provides distributed-memory matrix and vector, whose
data structures are presented by Fig. \ref{fig-ds-vec} and \ref{fig-ds-mat}. In Fig. \ref{fig-ds-vec},
vector information is stored, such as number of entries belong to current process (\verb|nlocal|)
and number of total entries (\verb|localsize|), including off-process entries. In Fig. \ref{fig-ds-mat},
matrix distribution information, communication pattern, MPI information, and global row and column indices
are stored.

\begin{figure}[!htb]
\centering
\begin{MyVerbatim}
typedef struct VEC_
{
    FLOAT    *data;
    INT      nlocal;
    INT      localsize;

} VEC;
\end{MyVerbatim}
\caption{Data structure of VEC}
\label{fig-ds-vec}
\end{figure}

\begin{figure}[!htb]
\centering
\begin{MyVerbatim}
/* struct for a matrix row */
typedef struct MAT_ROW_
{
    FLOAT   *data;
    INT     *cols;
    INT     *gcols;
    INT     ncols;

} MAT_ROW;

typedef struct MAT_
{
    /* data */
    MAT_ROW    *rows;

    MAP        *map;
    COMM_INFO  *cinfo;

    INT        nlocal;
    INT        localsize;
    INT        nglobal;

    int        rank;
    int        nprocs;
    MPI_Comm   comm;

} MAT;
\end{MyVerbatim}
\caption{Data structure of MAT}
\label{fig-ds-mat}
\end{figure}

Basic matrix-vector operations, such as
\begin{equation}
y = \alpha A x + \beta y,
\end{equation}
\begin{equation}
z = \alpha A x + \beta y,
\end{equation}
\begin{equation}
y = \alpha x + \beta y,
\end{equation}
\begin{equation}
z = \alpha x + \beta y,
\end{equation}
\begin{equation}
\alpha = \langle x, y \rangle,
\end{equation}
\begin{equation}
\alpha = \langle x, x \rangle^{\frac{1}{2}},
\end{equation}
are implemented.
With these operations, Krylov subspace solvers and preconditioners are implemented, including
GMRES, BiCGSTAB, Orthomin, RAS (Restricted Additive Schwarz) preconditioner and AMG preconditioner from HYPRE \citep{HYPRE2}.

\section{NUMERICAL EXPERIMENTS}
\label{sec-exp}

The systems are used for the numerical experiments. The first one is
an Blue Gene/Q from IBM.
The system, Wat2Q, is located in the IBM Thomas J. Watson Research Center.
Each node has 32 computer cards (64-bit PowerPC A2 processor), which has 17 cores.
One of them is for the operation system and the other
16 cores for computation. The system has 32,768 CPU cores for computation.
The performance of each core is really low
compared with Intel processors. However, the system has strong network relative to CPU performance,
and the system is scalable.
The second one is GPC from SciNet. It uses Intel Xeon E5540 CPU for computation and
InfiniBand for communication. Each node has two CPUs and the system has 3,864 nodes
(30,912 cores). The tests focus on scalability.

\subsection{Validation}
This section compares our results with commercial simulators and open results to check
the correctness of our implementation.

\begin{figure}[!htb]
\begin{center}
    \includegraphics[width=0.8\linewidth]{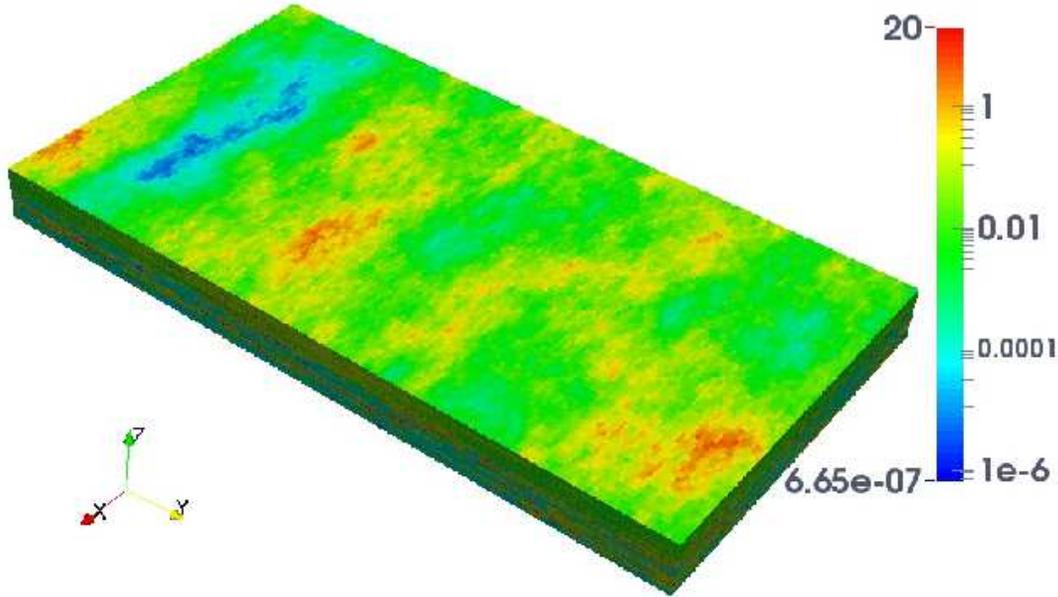}
\end{center}
\caption{Permeability in X Direction of the SPE10 benchmark}
\label{fig-spe10-perm}
\end{figure}

\begin{figure}[!htb]
\begin{center}
    \includegraphics[width=0.8\linewidth]{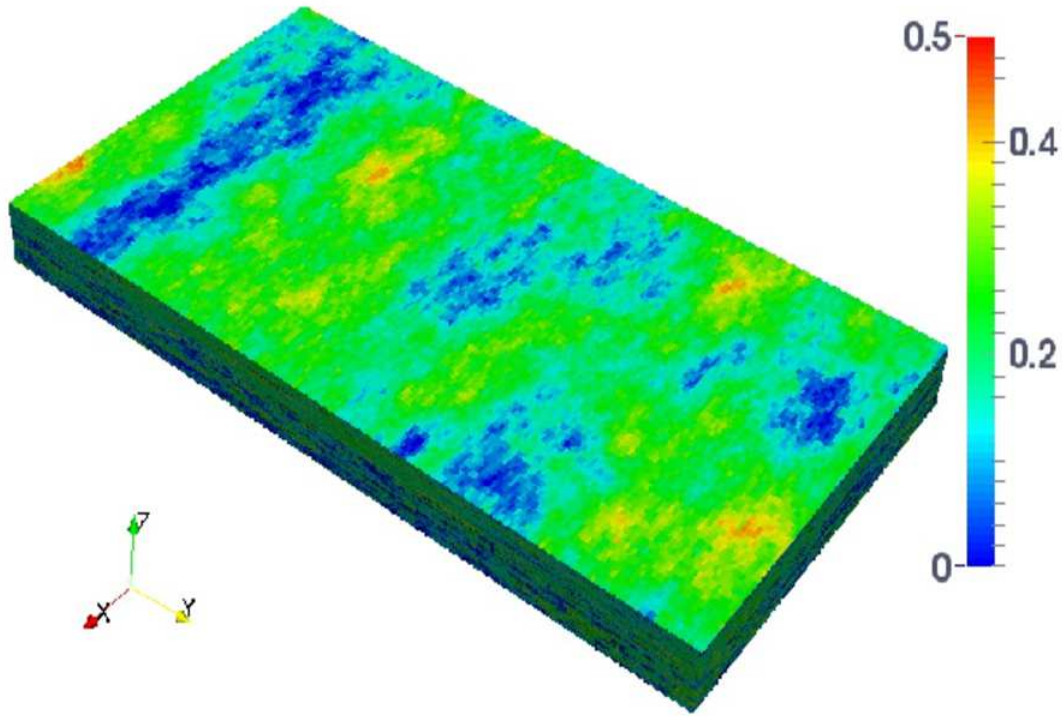}
\end{center}
\caption{Porosity of the SPE10 benchmark}
\label{fig-spe10-poro}
\end{figure}

\begin{figure}[!htb]
    \centering
    \includegraphics[width=0.5\linewidth, angle=270]{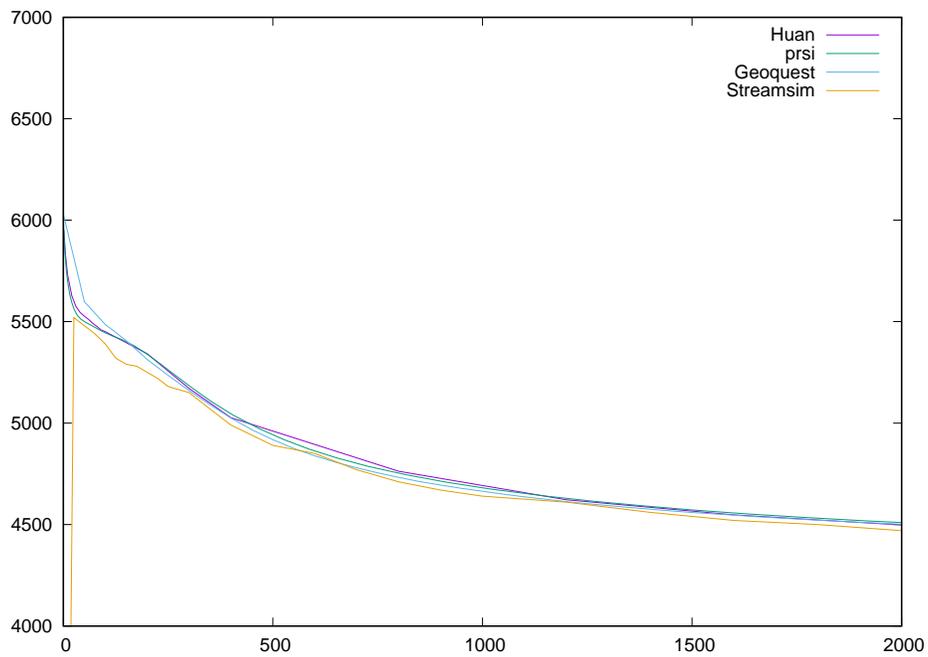}
    \caption{Average pressure of Example \ref{ex-spe10}}
    \label{fig-spe10-avg}
\end{figure}

\begin{figure}[!htb]
    \centering
    \includegraphics[width=0.5\linewidth, angle=270]{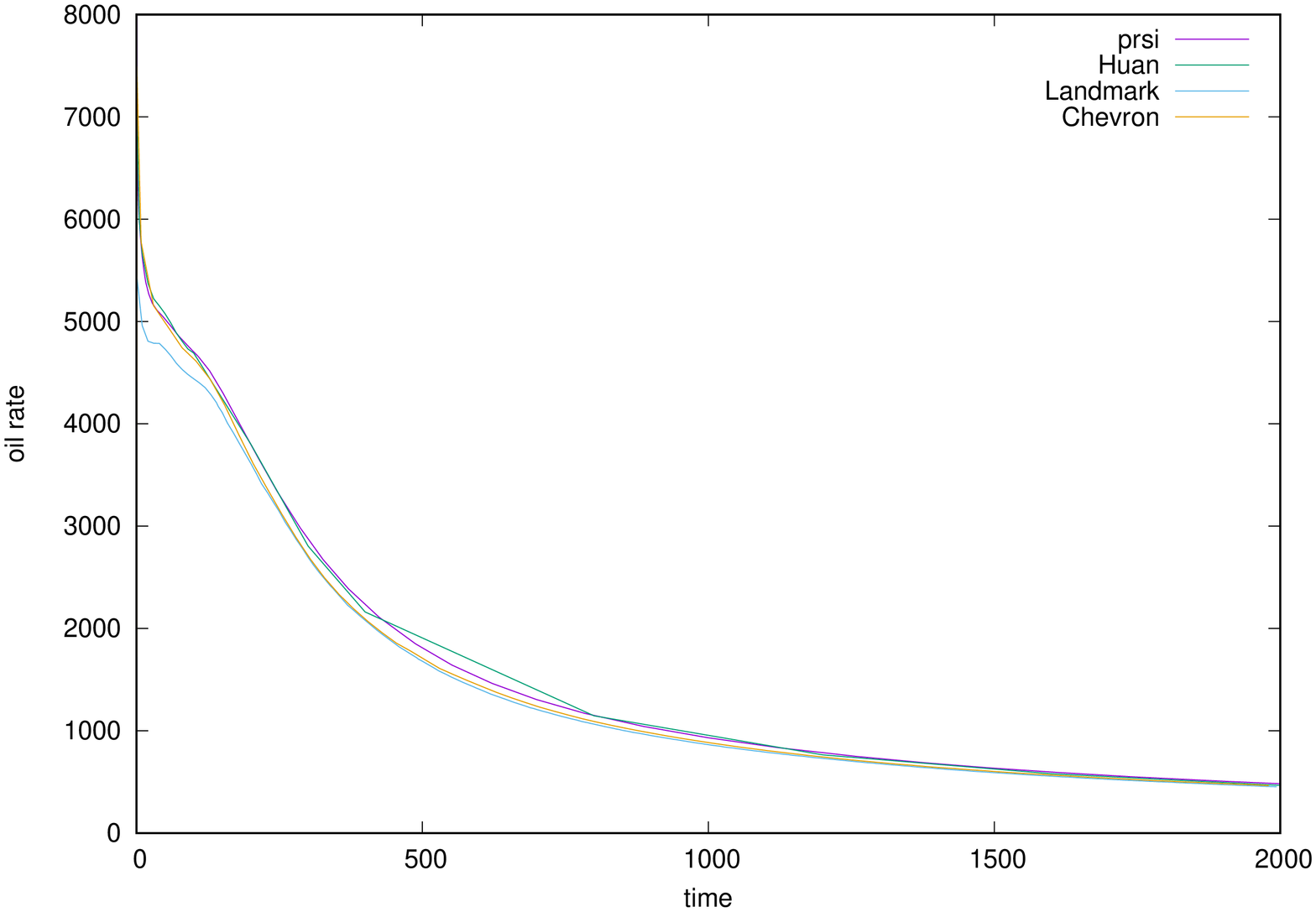}
    \caption{Oil rate of Example \ref{ex-spe10}}
    \label{fig-spe10-or}
\end{figure}

\begin{example}
\label{ex-spe10}
This example tests the Tenth SPE Comparative Solution Project, SPE10 \citep{SPE10},
an oil-water model, which has a sufficiently
fine grid. Its dimensions are
$1,200 \times 2,200 \times 170 $ (ft) and the fine scale cell size is $20 \times 10  \times 2$ (ft).
Its grid size is $60 \times 220 \times 85$ cells ($1.122 \times 10^6$ cells).
This model has five wells, one of them is injection well and four are production wells.
It has around 2.244 millions of unknowns.
The model is highly heterogeneous, whose permeability
is ranged from 6.65e-7 Darcy to 20 Darcy, and the x-direction permeability
is shown in Fig. \ref{fig-spe10-perm} \cite{bos-pc}. The porosity, which is demonstrated in Fig. \ref{fig-spe10-poro} \cite{bos-pc},
ranges from 0 to 0.5.
It relative permeability of water phase is calculated as
\begin{equation}
K_{rw} (s_w) = \frac{(s_w - s_{wc})^2}{(1 - s_{wc} - s_{or})^2},
\end{equation}
and the relative permeability of oil phase is calculated as
\begin{equation}
K_{ro} (s_w) = \frac{(1 - s_{or} - s_w)^2}{(1 - s_{wc} - s_{or})^2},
\end{equation}
where $s_{wc} = s_{or} = 0.2$.
More details can be found in the reference \citep{SPE10}.
The solver is GMRES(30) solver. The stopping tolerance for Newton method is 1e-3.
Our results are compared with other openly available results and presented in Fig. \ref{fig-spe10-avg}
and \ref{fig-spe10-or}.

\end{example}

Two measurements are reported, which are average pressure and oil production rate.
Fig. \ref{fig-spe10-avg} shows average pressure in each time step and it is compared with
results from commercial software, from which we can see that the results match
very well. Fig. \ref{fig-spe10-or} compares oil production rate with open
results from other companies.
Again, we can see these oil production results match.
This example indicates our implementation
is correct and our results match other simulators.

\begin{example}
The example test the standard black oil model and the data, mxspe009, is from CMG (Computer Modelling Group).
This model is heterogeneous with permeability varying from cell to cell,
porosity varying for each layer in $z$ direction.

The grid is 24$\times$ 25$\times$15 with mesh size 300ft. in $x$ and $y$ directions and
20, 15, 26, 15, 16, 14, 8, 8, 18, 12, 19, 18, 20, 50, 100 ft. in $z$ direction from top to bottom.
The depth of the top layer center is 9010 ft.
From top to bottom the porosity varies as 0.087, 0.097, 0.111, 0.16, 0.13,
0.17, 0.17, 0.08, 0.14, 0.13, 0.12, 0.105, 0.12, 0.116, 0.157.

The initial conditions are as follows:
bubble point pressure equals 3600.0 psi, reference pressure is 3600 psi at associated depth 9035 ft,
depth to water-oil contact is 9950 ft, depth to gas-oil contact is 8800 ft.

All wells are vertical. There is only one injection well with maximum water injection
rate 5000 bbl/day, maximum bottom hole pressure 4543.39 psi.
There are 25 production wells, maximum oil rate 1500 bbl/day.
More details can be found from CMG IMEX and \citep{spe9}.

\end{example}

\begin{figure}[!htb]
\includegraphics[scale = 0.65]{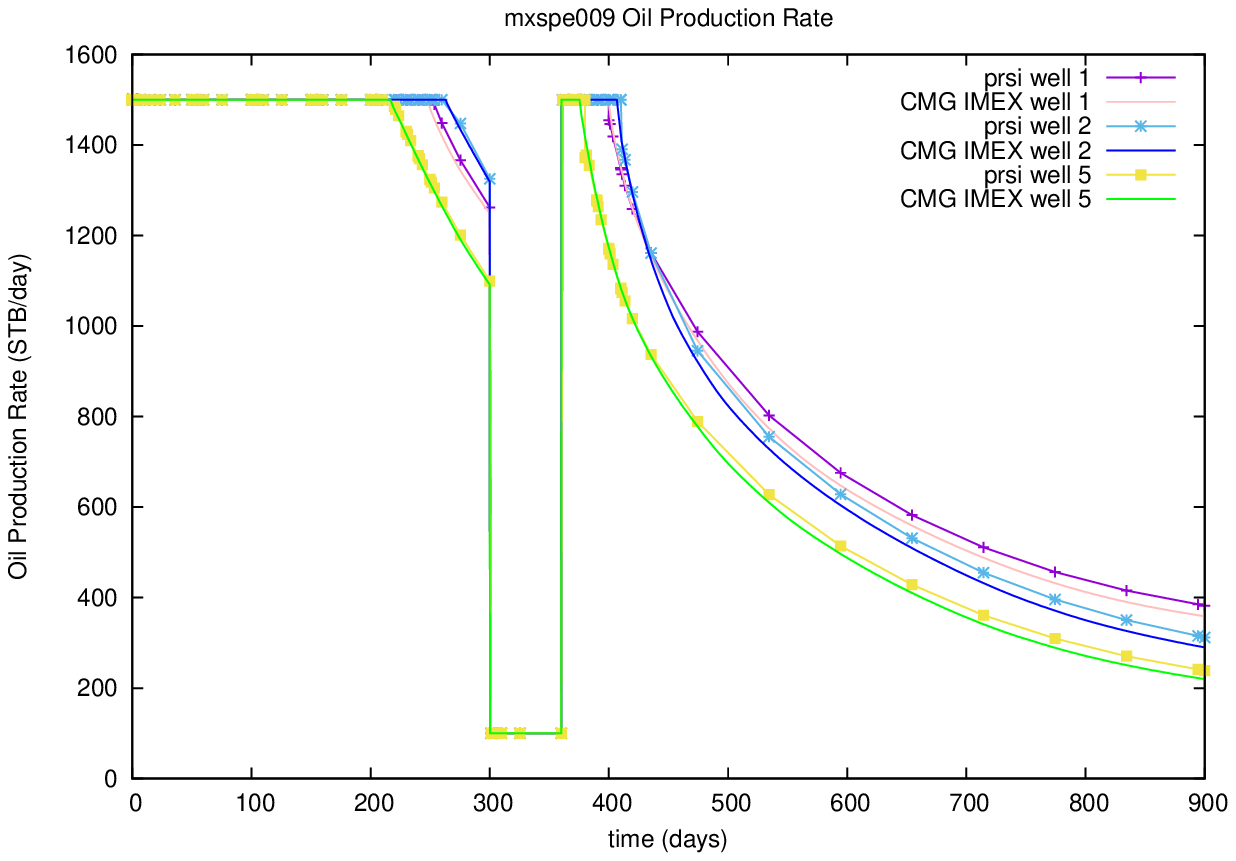}
\includegraphics[scale = 0.65]{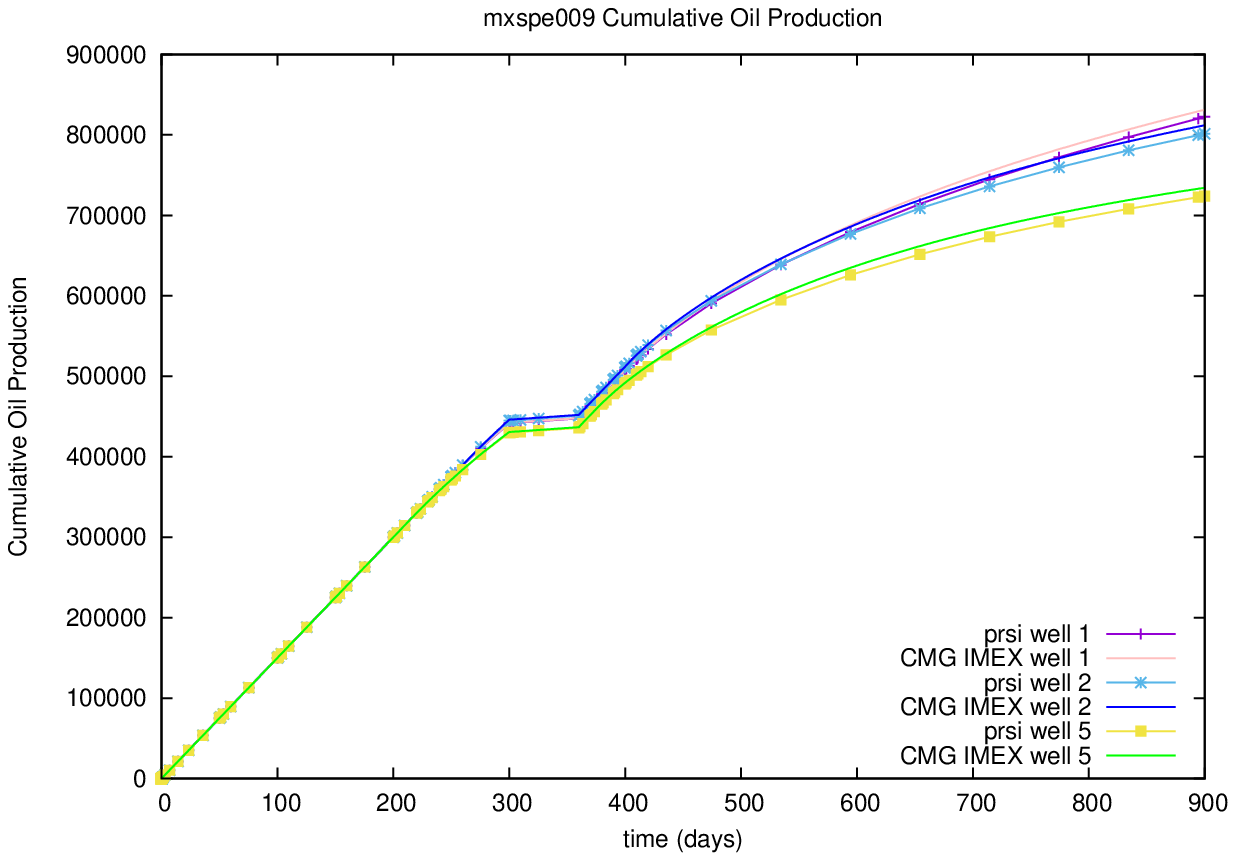}
\caption{Oil production rate of of well 1, 2, 5 and cumulative oil production (STB)
for mxspe009.}\label{spe9-oil}
\end{figure}

\begin{figure}[!htb]
\includegraphics[scale = 0.65]{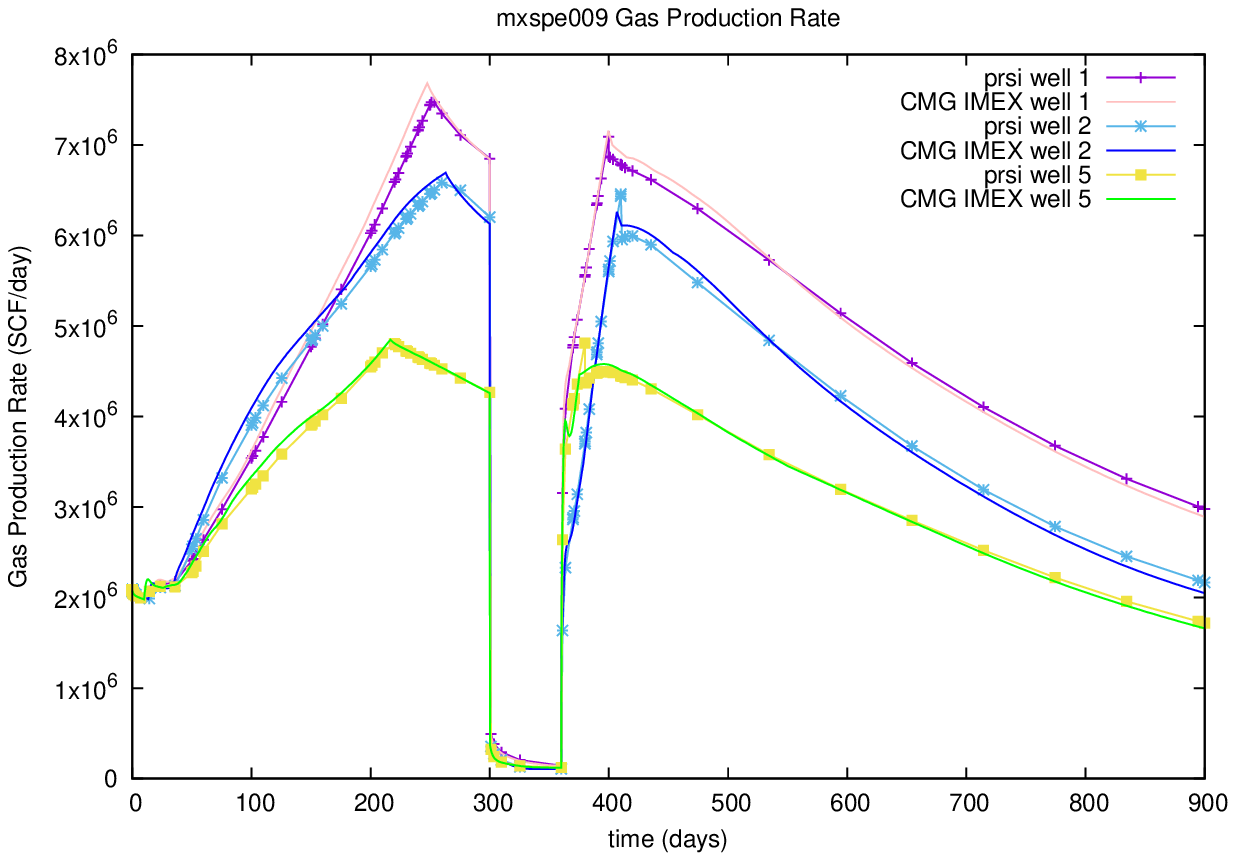}
\includegraphics[scale = 0.65]{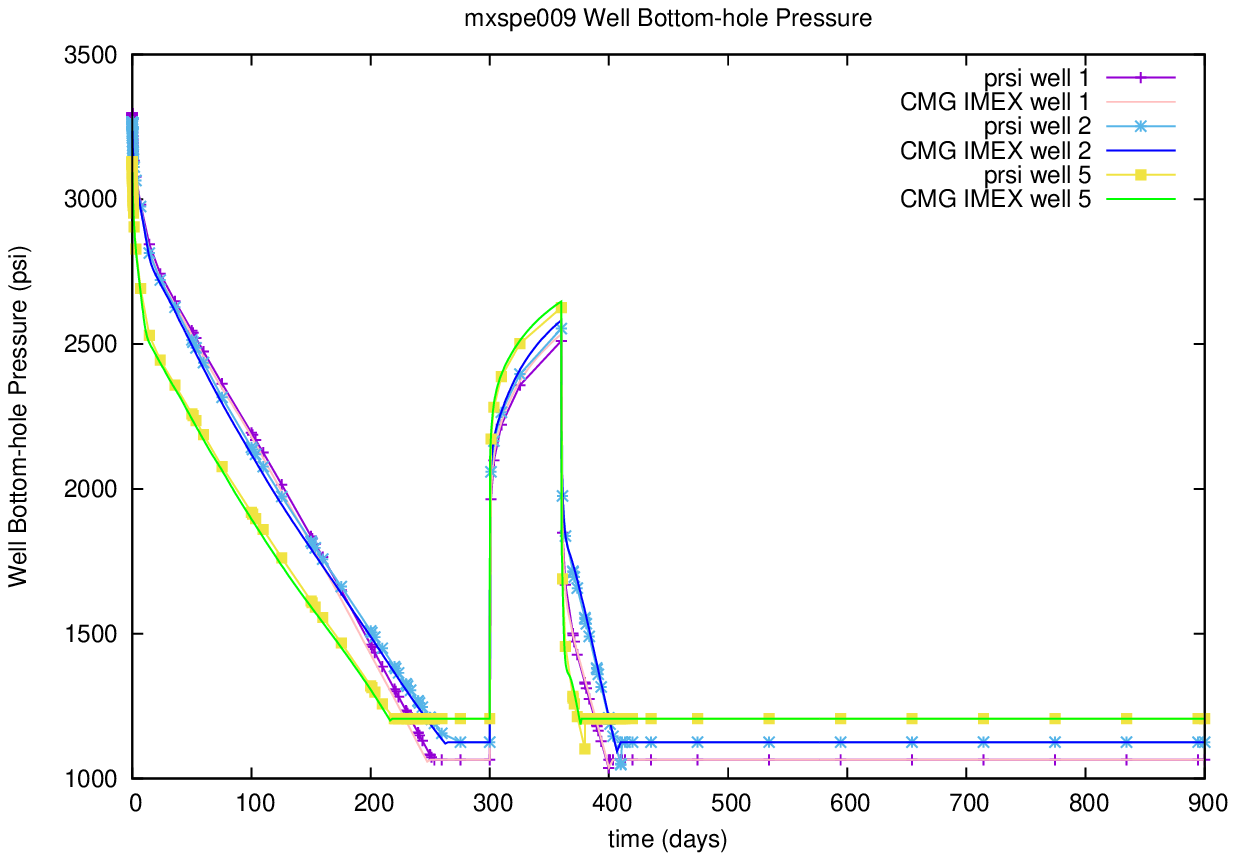}
\caption{Gas production rate and bottom-hole pressure of well 1, 2, 5 for mxspe009.}
\label{spe9-gas-rate-bhp}
\end{figure}

The results from CMG IMEX are marked with "CMG IMEX" and the results
from our simulator are marked with "prsi".
Fig. \ref{spe9-oil} shows the oil production rates of production well 1, 2, 5 and their
cumulative oil production.
Fig. \ref{spe9-gas-rate-bhp} shows the gas production rates and bottom-hole pressures of well 1, 2, 5.
From the two figures, we can see that the results of our simulator match results from CMG IMEX.

\begin{example}

This is a model of two-phase oil-water problem in naturally fractured reservoir.
The dual porosity dual permeability method is applied. The model, mxfrr003, is from CMG IMEX.
The grid is 10$\times$ 10$\times$1 with mesh size 102.04ft. in $x$ and $y$ directions
and 100.00ft. in $z$ direction.
The depth of the top layer center is 9010 ft.

The porosities of the matrix and the fracture are 0.1392 and 0.039585, respectively.
The permeability for the matrix is 100mD in $x$ and $y$ directions and 10mD in $z$ direction.
The permeability for the fracture is 450md, which is 395.85mD in the original model.
A new relative permeability curve for water phase is applied to fracture.

The initial conditions are as follows:
the bubble point pressure equals 15.0 psi for both matrix and fracture, the initial pressure 1479.0 psi
and 1463.0 psi for matrix and fracture respectively.
The oil saturation for matrix and fracture 0.92 and 0.99, respectively.

There is one injection well and one production well, both of which are vertical wells.
The injection well has maximum 500 bbl/day water injection rate.
The perforation is at first cell.
The production well has maximum 500 bbl/day liquid production rate and minimum 15 psi bottom-hole pressure.
The total simulation time is 1600 days.

\end{example}
\begin{figure}[!htb]
    \centering
\includegraphics[scale = 0.65]{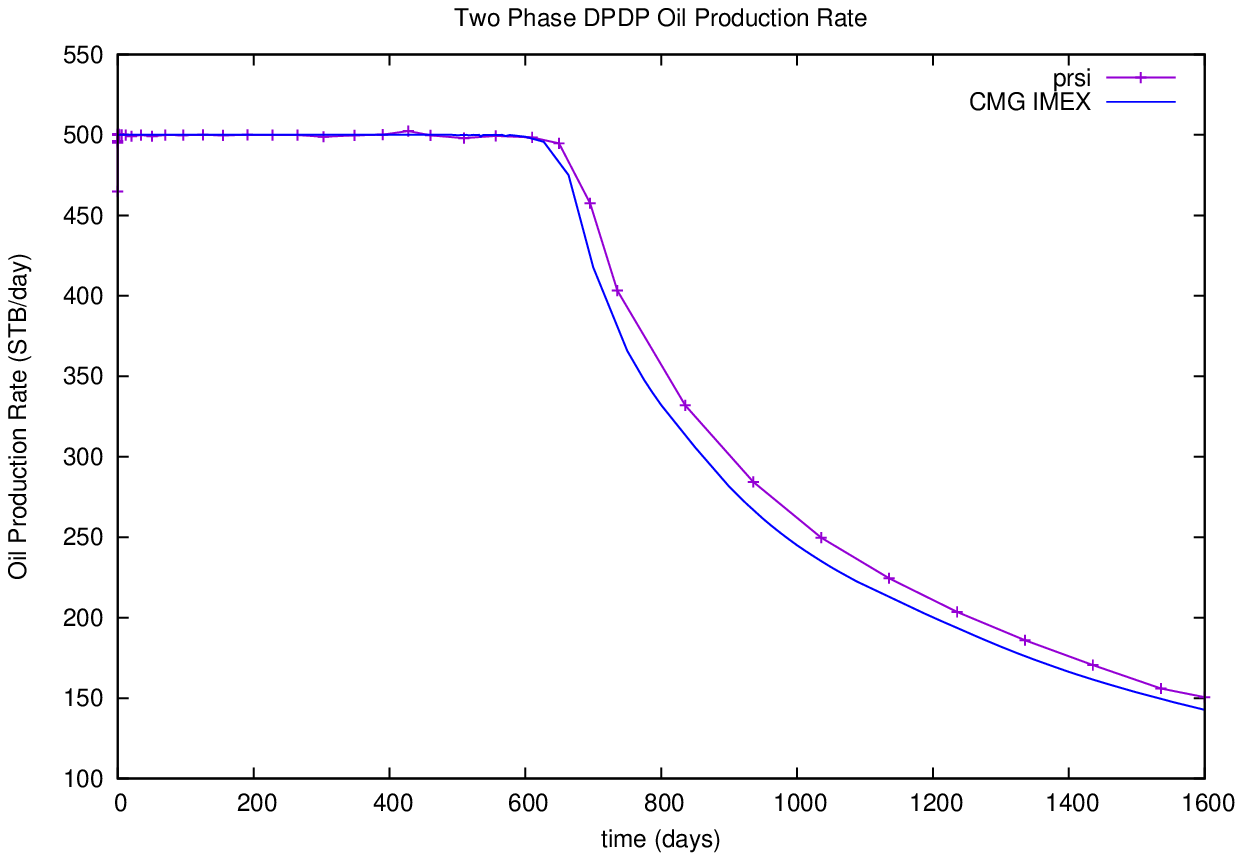}
\includegraphics[scale = 0.65]{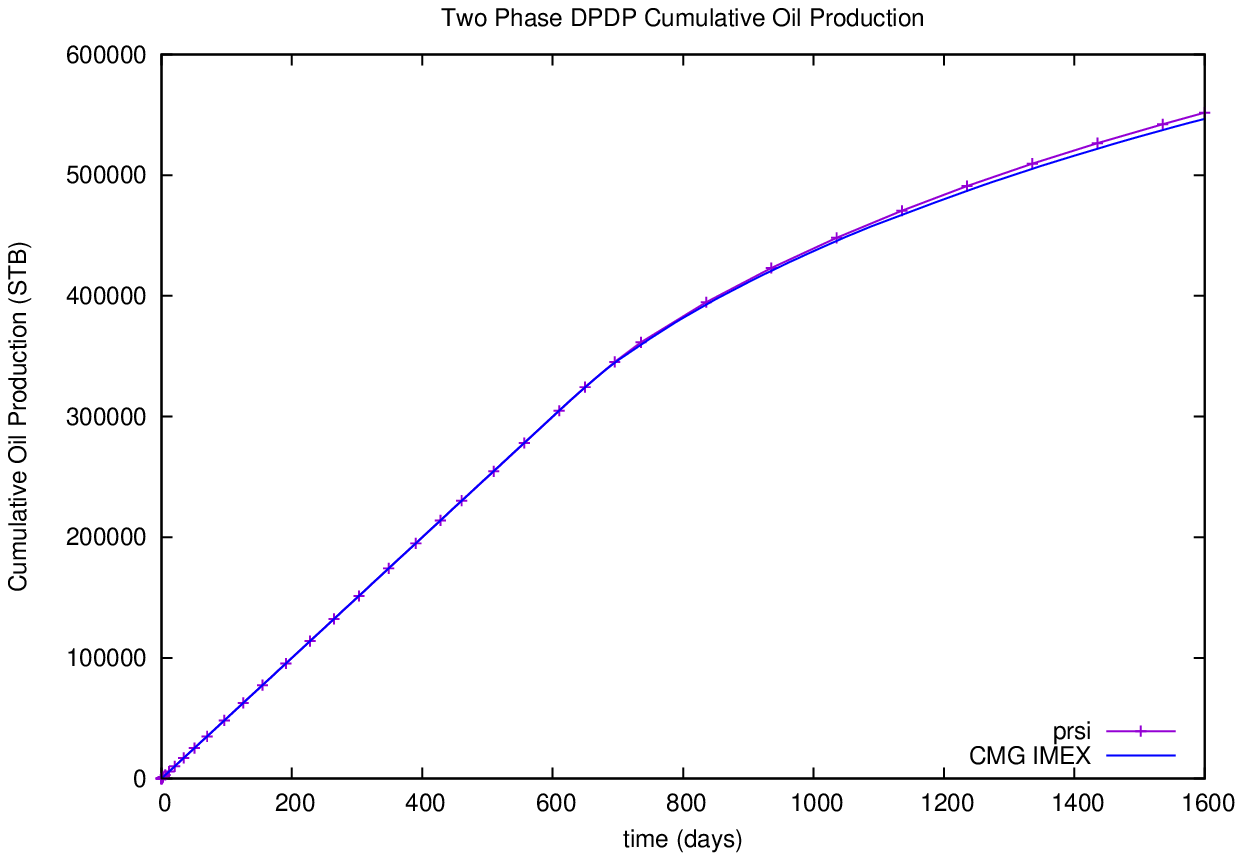}
\caption{Oil production rate and cumulative oil production of injector and producer for mxfrr003.}\label{frr3-oil}
\end{figure}

\begin{figure}[!htb]
\centering
\includegraphics[scale = 0.75]{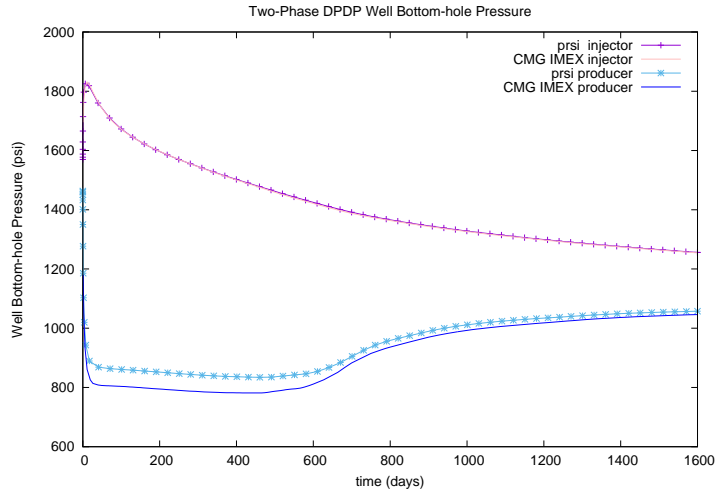}
\caption{Bottom-hole pressure of injection and production wells for mxfrr003.}
\label{frr3-bhp}
\end{figure}

\begin{figure}[!htb]
\centering
\includegraphics[scale = 0.65]{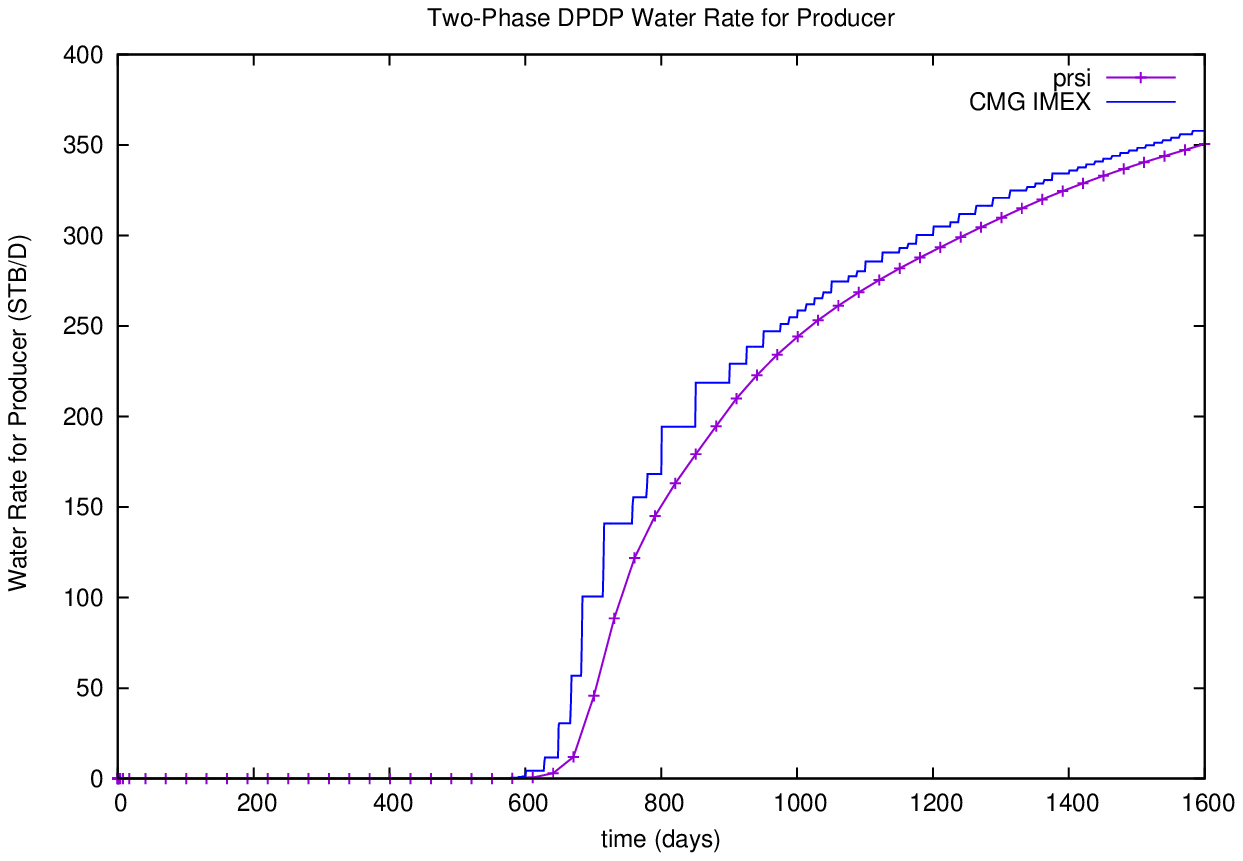}
\includegraphics[scale = 0.65]{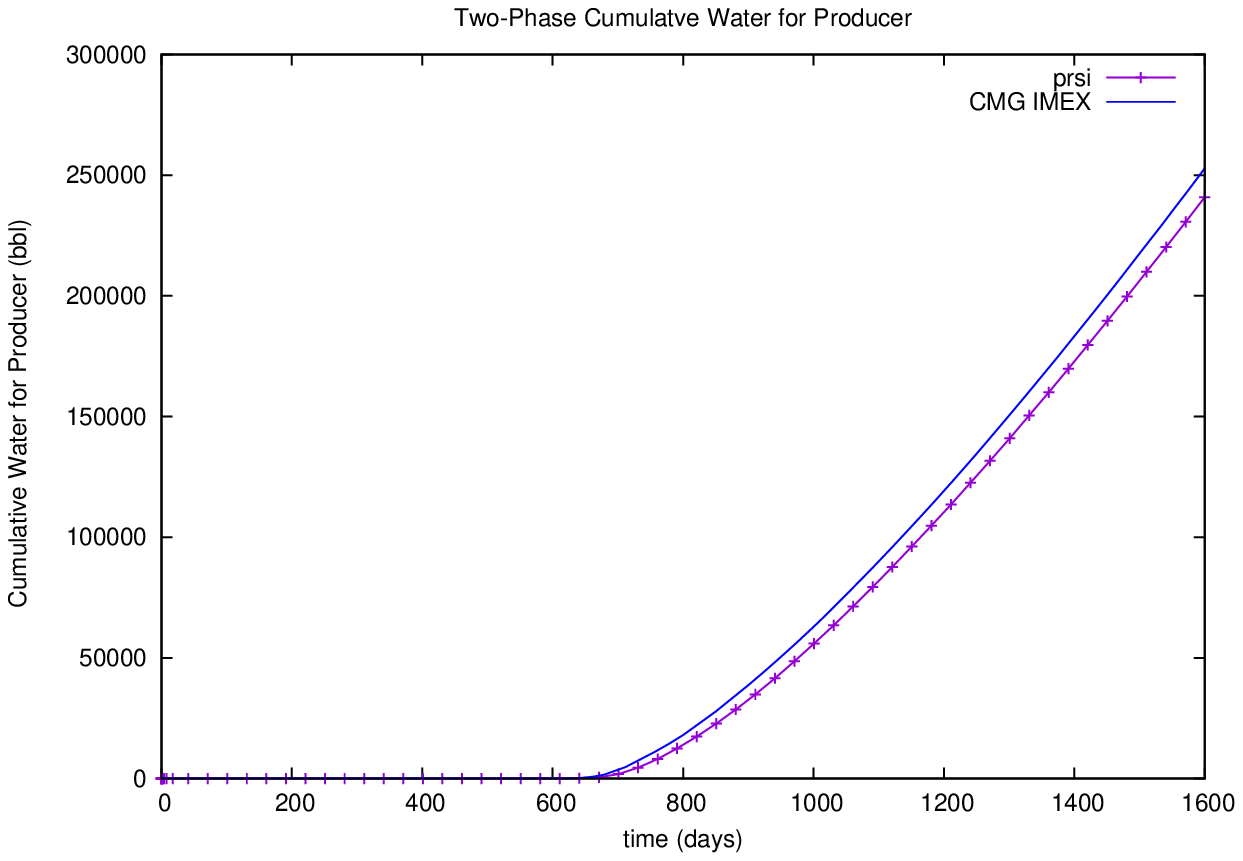}
\caption{Water rate and cumulative production of production well for mxfrr003.}
\label{frr3-wr}
\end{figure}

\begin{figure}[!htb]
\centering
\includegraphics[scale = 0.75]{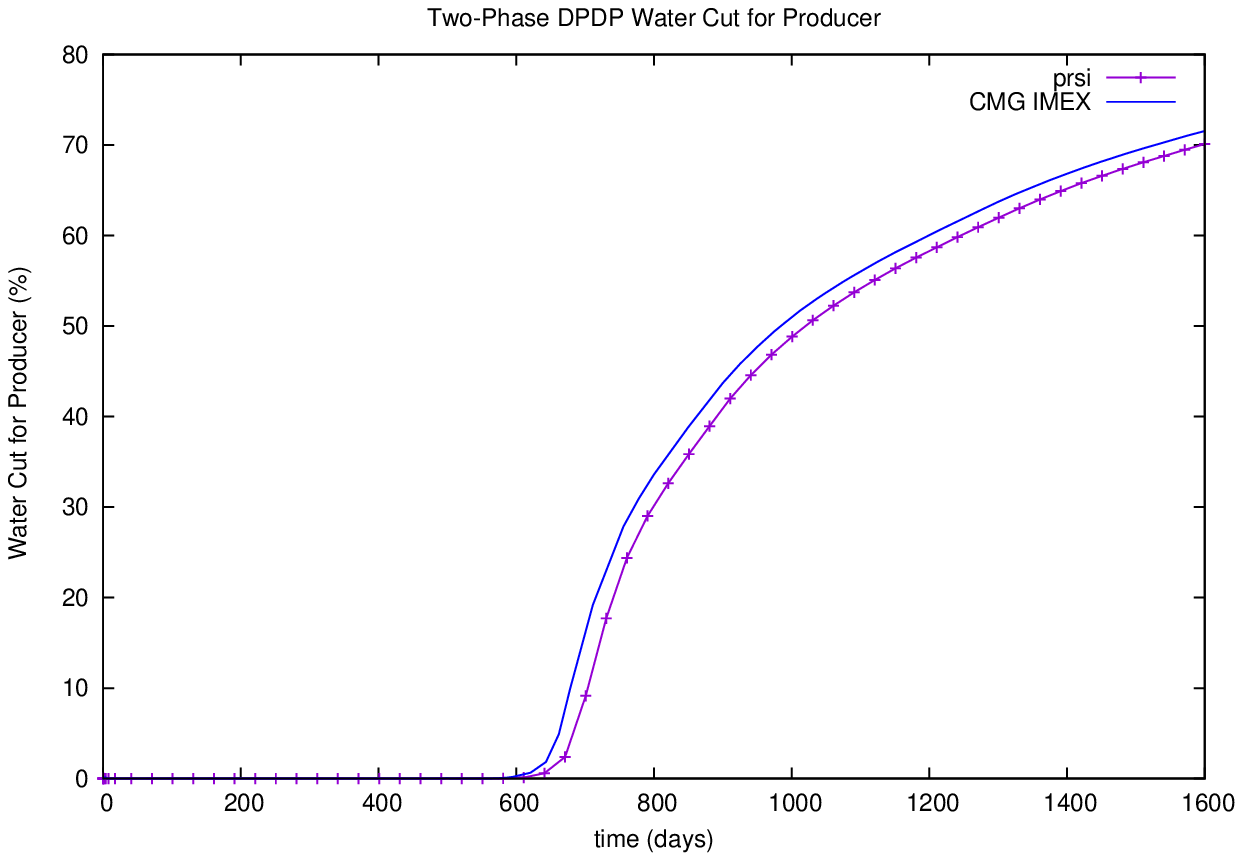}
\caption{Water cut of production well for mxfrr003.}
\label{frr3-wc}
\end{figure}

The oil production rates and cumulative oil production are shown in Fig. \ref{frr3-oil}. Again, results from
CMG IMEX are marked with "CMG IMEX" and results from our simulator are marked with "prsi".
The bottom-hole pressures of injector and producer are shown in Fig. \ref{frr3-bhp}.
Water production rate, cumulative production and water cut are shown in Fig. \ref{frr3-wr}
and Fig. \ref{frr3-wc}, respectively. We can see that
our results match CMG IMEX's results, which indicates that our implementation is correct.


\subsection{Scalability}
This section presents scalability results using two parallel systems.

\begin{figure}[!htb]
    \centering
    \includegraphics[width=0.5\linewidth, angle=270]{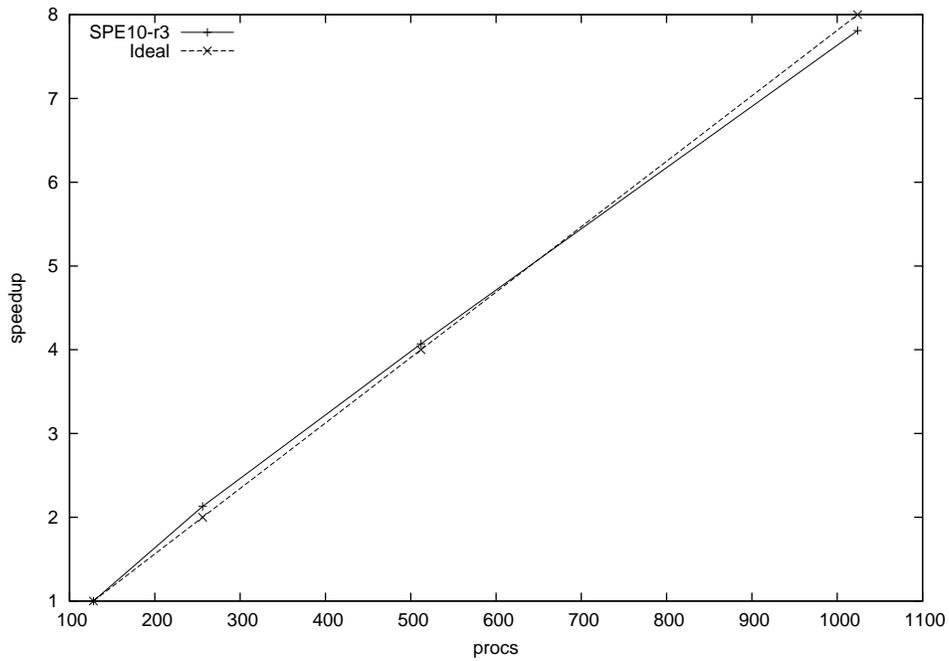}
    \caption{Scalability of Example \ref{ow-r3-bgq}}
    \label{fig-ow-r3-bgq}
\end{figure}

\begin{example}
\label{ow-r3-bgq}
The case tests two-phase oil-water model with a
refined SPE10 project and each original cell is refined to 27 smaller cells. The model
has around 30 millions of grid cells. The stopping criteria
for inexact Newton method is 1e-2 and maximal nonlinear iterations are 20.
The linear solver is BiCGSTAB, and its maximal iterations are 100. The simulation time is
10 days. The case is run on IBM Blue Gene/Q. Numerical summaries are shown in Table \ref{tab-ow-r3-bgq}
and scalability is presented in Fig \ref{fig-ow-r3-bgq}.
\end{example}

\begin{table}[!htb]
\centering
  \caption{Numerical summaries of Example \ref{ow-r3-bgq}}
\begin{tabular}{ccccc} \\ \hline
  \# procs   & \# Steps & \# Newton & \# Solver  & Time (s)\\ \hline
128 & 40 & 295 & 2470 & 43591.87 \\
256 & 39 & 269 & 2386 & 20478.49 \\
512 & 40 & 260 & 2664 & 10709.86 \\
1024 & 39 & 259 & 2665 & 5578.75 \\
 \hline
\end{tabular}
  \label{tab-ow-r3-bgq}
\end{table}

This case uses up to 1024 MPI processes and their speedups are compared with case that uses
128 MPI tasks. From Table \ref{tab-ow-r3-bgq}, we can see Newton method and linear solver are robust.
When more MPI tasks are employed, fewer Newton iterations are required. And each Newton iteration
terminates in around 10 linear iterations. The preconditioner is effective.
The running time and Fig. \ref{fig-ow-r3-bgq}
show the parallel simulator has excellent scalability, which is almost ideal on IBM Blue Gene/Q.
The results also show that our solver and preconditioner are scalable.

\begin{figure}[!htb]
    \centering
    \includegraphics[width=0.5\linewidth, angle=270]{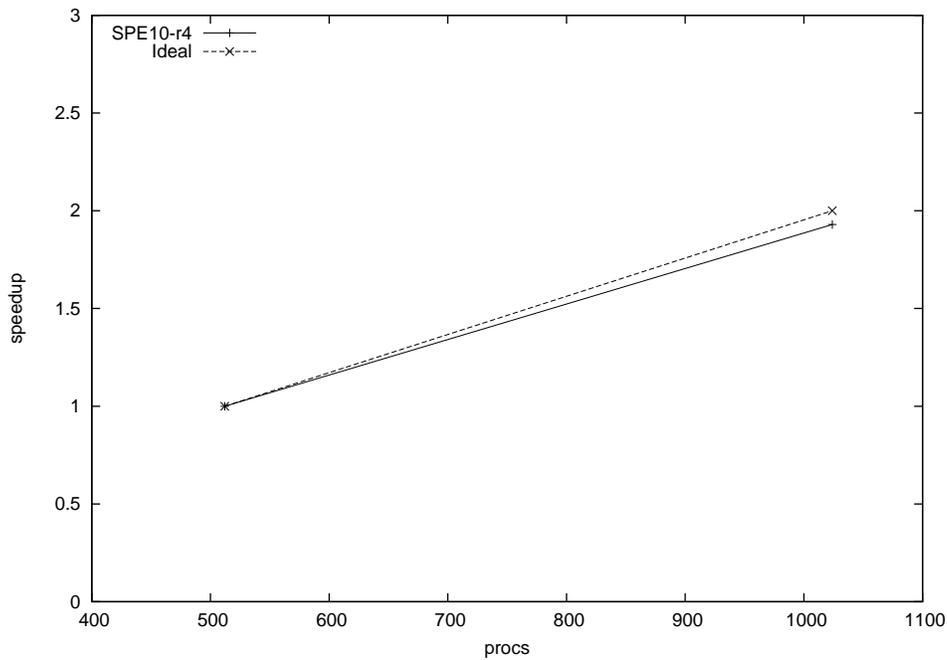}
    \caption{Scalability of Example \ref{ow-r4-gpc}}
    \label{fig-ow-r4-gpc}
\end{figure}

\begin{example}
\label{ow-r4-gpc}
The case also tests two-phase oil-water model, where
a refined SPE10 project is used and each cell is refined to 64 smaller cells. It
has around 65 millions of grid cells. The stopping criteria
for inexact Newton method is 1e-2 and its maximal nonlinear iterations are 20.
The linear solver is BiCGSTAB, and its maximal iterations are 100. The simulation time is
20 days. The case is run on GPC (General Purpose Cluster). Numerical summaries are shown in Table \ref{tab-ow-r4-gpc}
and scalability is presented in Fig. \ref{fig-ow-r4-gpc} \citep{lrs2016}.
\end{example}

\begin{table}[!htb]
\centering
  \caption{Numerical summaries of Example \ref{ow-r4-gpc}}
\begin{tabular}{ccccc} \\ \hline
  \# procs   & \# Steps & \# Newton & \# Solver  & Time (s)\\ \hline
  512  & 107 & 662 & 6971 & 26636.13  \\
  1024 & 108 & 668 & 7427 & 13772.96  \\ \hline
\end{tabular}
  \label{tab-ow-r4-gpc}
\end{table}

This case is larger than last example and is run on a different parallel system. Two different
configurations are benchmarked. Again, Table \ref{tab-ow-r4-gpc} shows our nonlinear method
and linear solver are robust. Each time step uses around 6.5 Newton iterations. Our linear solver
and preconditioner are effective, which can solve a linear system in around 11 linear iterations.
The scalability is demonstrated by Fig. \ref{fig-ow-r4-gpc}.

\begin{figure}[!htb]
    \centering
    \includegraphics[width=0.5\linewidth, angle=270]{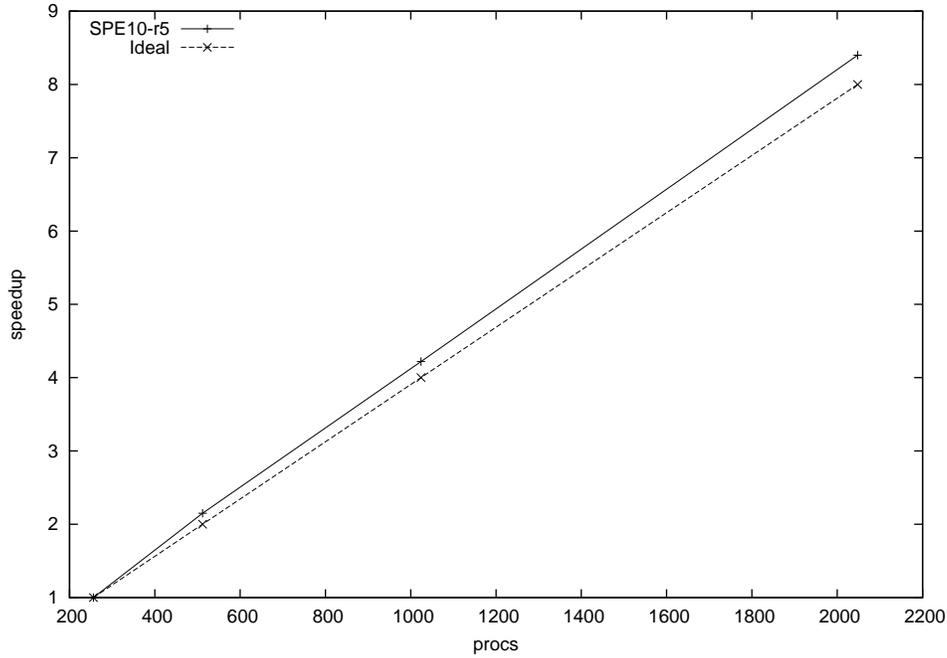}
    \caption{Scalability of Example \ref{ow-r5-bgq}}
    \label{fig-ow-r5-bgq}
\end{figure}

\begin{example}
\label{ow-r5-bgq}
The case tests two-phase oil-water model with a
refined SPE10 project and each cell is refined to 125 cells. The model
has around 140 millions of cells. The stopping criteria
for inexact Newton method is 1e-2 and maximal nonlinear iterations are 20.
The linear solver is BiCGSTAB, and its maximal iterations are 50. The simulation time is
10 days. The case is run on IBM Blue Gene/Q.
Numerical summaries are shown in Table \ref{tab-ow-r5-bgq}
and scalability is presented in Fig. \ref{fig-ow-r5-bgq}.
\end{example}

\begin{table}[!htb]
\centering
  \caption{Numerical summaries of Example \ref{ow-r5-bgq}}
\begin{tabular}{ccccc} \\ \hline
  \# procs   & \# Steps & \# Newton & \# Solver  & Time (s)\\ \hline
256 & 27 & 108 & 495 & 41127.23 \\
512 & 27 & 105 & 515 & 19112.77 \\
1024 & 27 & 102 & 572 & 9756.6 \\
2048 & 26 & 101 & 625 & 4896.47 \\
 \hline
\end{tabular}
  \label{tab-ow-r5-bgq}
\end{table}

From Table \ref{tab-ow-r5-bgq}, we can see Newton method is robust. The linear solver
is also robust. However, when more MPI tasks are employed, its convergence becomes lower
and lower. Average of linear iterations increases from 4.9 to 6.2. Even through,
Fig. \ref{fig-ow-r5-bgq} show the simulator has linear scalability.

\begin{figure}[!htb]
    \centering
    \includegraphics[width=0.6\linewidth, angle=270]{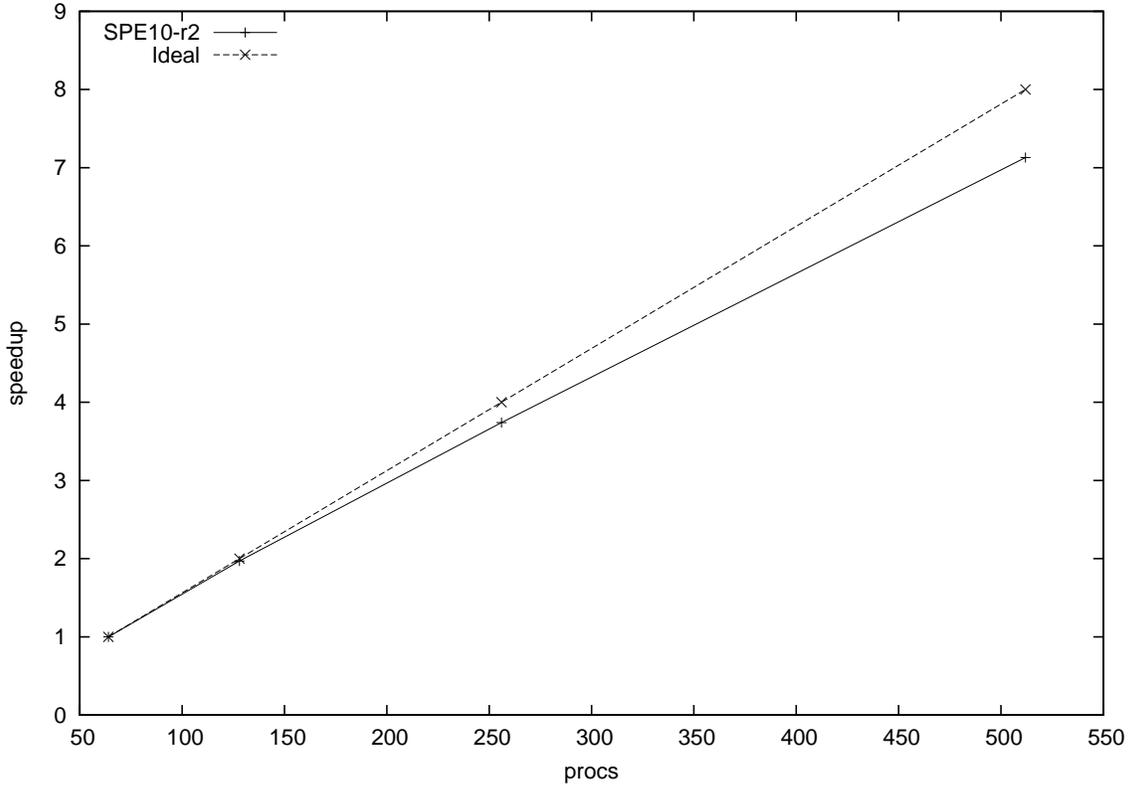}
    \caption{Scalability of Example \ref{bos-r2-gpc}}
    \label{fig-bos-r2-gpc}
\end{figure}

\begin{example}
\label{bos-r2-gpc}
This case tests the standard black oil model with a refined SPE10
geological model and each cell is refined to 8 cells. It has around 8.96 millions
of grid cells. The stopping criteria
for inexact Newton method is 1e-3 and maximal nonlinear iterations are 15.
The linear solver is BiCGSTAB, and its maximal iterations are 100. The simulation time is
200 days. The case is run on GPC (General Purpose Cluster). Numerical summaries are shown in Table \ref{tab-bos-r2-gpc}
and scalability is presented in Fig. \ref{fig-bos-r2-gpc}.
\end{example}

\begin{table}[!htb]
\centering
  \caption{Numerical summaries of Example \ref{bos-r2-gpc}}
\begin{tabular}{ccccc} \\ \hline
  \# procs   & \# Steps & \# Newton & \# Solver  & Time (s)\\ \hline
  64 & 219 & 1444 & 23597 & 82305.88 \\
  128 & 214 & 1402 & 23355 & 41859.71 \\
  256 & 218 & 1453 & 26934 & 22024.53 \\
  512  & 214 & 1401 & 24579 & 11548 \\ \hline
\end{tabular}
  \label{tab-bos-r2-gpc}
\end{table}

This example tests the standard black oil model. Table \ref{tab-bos-r2-gpc}
shows Newton method and linear solver are robust. The running time and
Fig. \ref{fig-bos-r2-gpc} show the simulator on GPC has a linear scalability.

\begin{figure}[!htb]
    \centering
    \includegraphics[width=0.5\linewidth, angle=270]{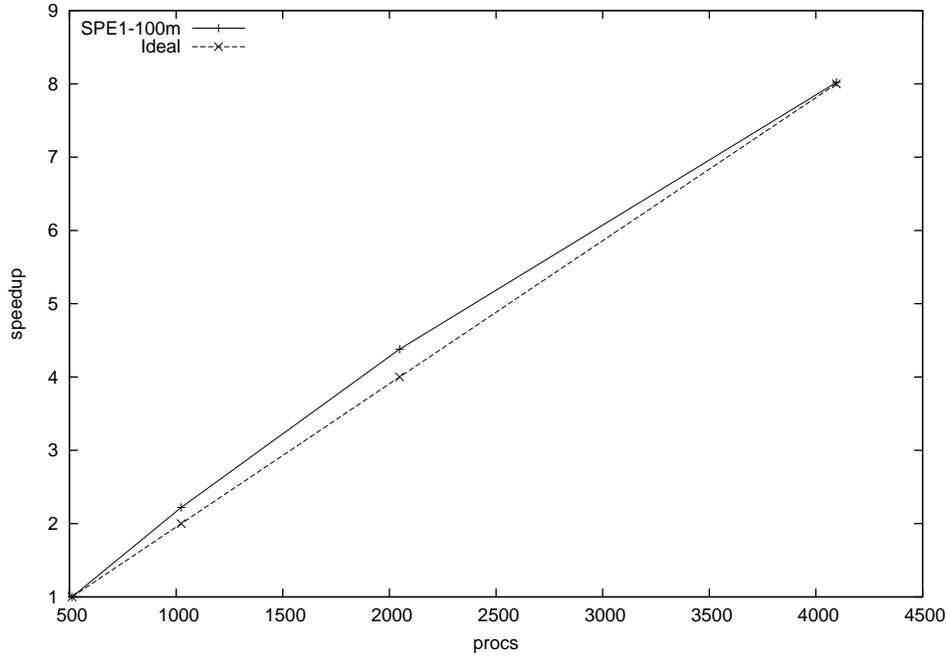}
    \caption{Scalability of Example \ref{bos-spe1-bgq}}
    \label{fig-bos-spe1-bgq}
\end{figure}

\begin{example}
\label{bos-spe1-bgq}
The case tests the standard black oil model using a refined SPE1 project.
The model has 100 millions of cells. The stopping criteria
for inexact Newton method is 1e-2 and maximal nonlinear iterations are 15.
The linear solver is BiCGSTAB, and its maximal iterations are 20. The simulation time is
10 days. The case is run on IBM Blue Gene/Q. Numerical summaries are shown in Table \ref{tab-bos-spe1-bgq}
and scalability is presented in Fig \ref{fig-bos-spe1-bgq}.
\end{example}

\begin{table}[!htb]
\centering
  \caption{Numerical summaries of Example \ref{bos-spe1-bgq}}
\begin{tabular}{ccccc} \\ \hline
  \# procs   & \# Steps & \# Newton & \# Solver  & Time (s)\\ \hline
512 & 27 & 140 & 586 & 11827.99 \\
1024 & 27 & 129 & 377 & 5328.46 \\
2048 & 26 & 122 & 362 & 2703.51 \\
4096 & 27 & 129 & 394 & 1474.21 \\
 \hline
\end{tabular}
  \label{tab-bos-spe1-bgq}
\end{table}

The case with 512 MPI tasks is the base case. From Table \ref{tab-bos-spe1-bgq}, we
can see Newton method and linear solver show good convergence. Fig. \ref{fig-bos-spe1-bgq}
shows the simulator has excellent scalability and cases with 1024 MPI tasks and 2048 MPI
tasks have super-linear scalability.

\begin{figure}
    \centering
    \includegraphics[width=0.5\linewidth, angle=270]{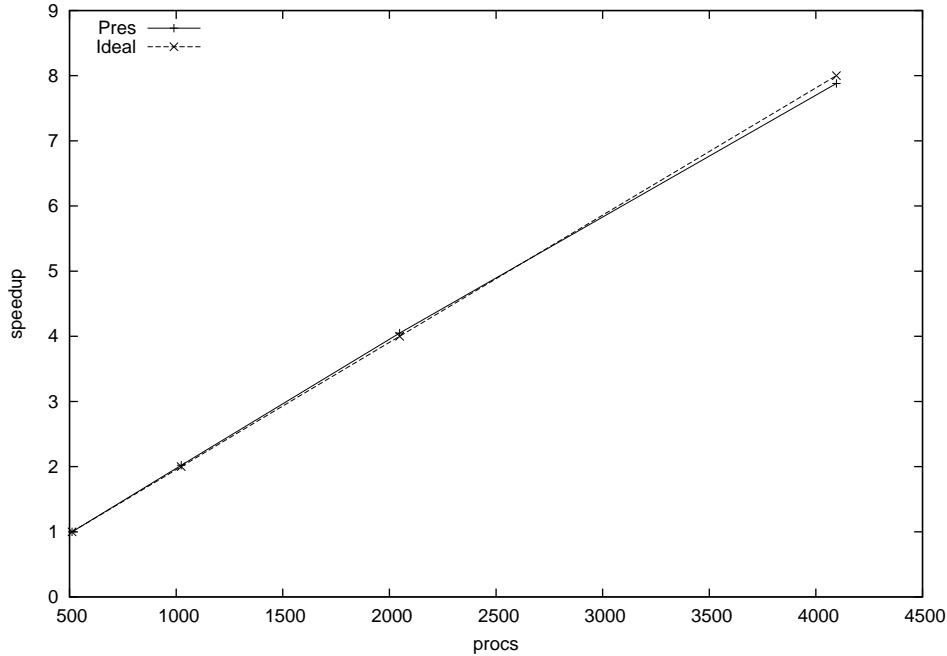}
    \caption{Scalability of Example \ref{po-bgq}}
    \label{fig-po-bgq}
\end{figure}

\begin{example}
\label{po-bgq}
The case tests one linear system from pressure equation and the size of the
matrix is 3 billion. GMRES(30) solver is applied and it has fixed iterations of 90.
The preconditioner is the RAS (Restricted Additive Schwarz) method.
The case is run on IBM Blue Gene/Q. Numerical summaries are shown in Table \ref{tab-po-bgq}
and scalability is presented in Fig \ref{fig-po-bgq}.
\end{example}

\begin{table}[!htb]
\centering
  \caption{Numerical summaries of Example \ref{po-bgq}}
\begin{tabular}{ccc} \\ \hline
  \# procs  & \# Solver  & Time (s) \\ \hline
  512 & 90 & 918.91 \\
  1024 & 90 & 454.04 \\
  2048& 90 & 227.05 \\
  4096 & 90 & 116.63 \\
 \hline
\end{tabular}
  \label{tab-po-bgq}
\end{table}

This example tests the scalability of linear solver, preconditioner and SpMV. Table \ref{tab-po-bgq}
shows that when MPI tasks are doubled, running time is cut by half. The numerical results and
Fig. \ref{fig-po-bgq} demonstrate the simulator can model extremely large-scale reservoirs and
it has excellent scalability.

\section{CONCLUSION}
A parallel reservoir simulator is presented, which can calculate standard black oil model and
oil-water model in
regular reservoirs and naturally fractured reservoirs. Their mathematical models, numerical methods
and parallel implementation are introduced.
Numerical experiments show that results from our simulator match results from other simulators and
our simulator has excellent scalability.
The paper also demonstrates that parallel computing is a powerful tool
for large-scale reservoir simulations.

\section*{ACKNOWLEDGEMENTS}
The support of Department of Chemical and Petroleum Engineering,
University of Calgary and Reservoir Simulation  Research Group is gratefully
acknowledged. The research is partly supported by NSERC, AIEES, Foundation
CMG, AITF iCore, IBM Thomas J. Watson Research Center, the Frank
and Sarah Meyer FCMG Collaboration Centre for Visualization and Simulation,
 WestGrid (www.westgrid.ca), SciNet (www.scinethpc.ca)
and Compute Canada (www.computecanada.ca).


\bibliographystyle{latex8}

\begin{thebibliography}{00}

\bibitem[Rutleedge et al., 1991]{PS-Rut}
J. Rutledge, D. Jones, W. Chen, and E. Chung, The Use of Massively Parallel SIMD
  Computer for Reservoir Simulation, SPE Computer Applications, 1991, 04, 16.

\bibitem[Shiralkar et al., 1997]{PS-Shi}
G. Shiralkar, R. Stephenson, W. Joubert, O. Lubeck, and B. van Bloemen Waanders,
A production quality distributed memory reservoir simulator,
SPE Reservoir Evaluation \& Engineering, 1998, 1, 400.

\bibitem[Kaarstad et al., 1995]{PS-Kaa}
T. Kaarstad, J. Froyen, P. Bjorstad, and M. Espedal, 
A Massively Parallel Reservoir Simulator,
Symposium on Reservoir Simulation, Society of Petroleum Engineers,
San Antonio, 12-15 February 1995.

\bibitem[Killough et al., 1997]{PS-Kil2}
J. Killough, D. Camilleri, B. Darlow, and J. Foster,
Parallel Reservoir Simulator Based on Local
Grid Refinement, SPE-37978, SPE Reservoir
Simulation Symposium, Dallas, 1997.

\bibitem[Dogru et al., 2009]{PS-Dogru2}
A. Dogru, L. Fung, U. Middya, T. Al-Shaalan, and J. Pita,
A next-generation parallel reservoir simulator for giant reservoirs,
SPE/EAGE Reservoir Characterization \& Simulation Conference. 2009.

\bibitem[Zhang, 2009]{phg}
{L. Zhang}, {A Parallel Algorithm for Adaptive Local Refinement of Tetrahedral Meshes
Using Bisection}, Numer. Math.: Theory, Methods and Applications, 2009, 2, 65--89.

\bibitem[Zhang et al., 2009]{phg-quad}
{L. Zhang, T. Cui, and H. Liu}, {A set of symmetric quadrature rules on triangles and tetrahedra},
J. Comput. Math, 2009, 27(1), 89--96.

\bibitem[Wallis et al., 1985]{CPR-old}
J. Wallis, R. Kendall, and T. Little,
Constrained residual acceleration of conjugate residual methods,
SPE Reservoir Simulation Symposium, 1985.

\bibitem[Cao et al., 2005]{CPR-cao}
H. Cao, T. Schlumberger, A. Hamdi, J. Wallis, and H. Yardumian,
Parallel scalable unstructured CPR-type linear solver for reservoir simulation.
SPE Annual Technical Conference and Exhibition. 2005.

\bibitem[Al-Shaalan et al., 2009]{Study-Two-Stage}
T. Al-Shaalan, H. Klie, A. Dogru, and M. Wheeler,
Studies of Robust Two Stage Preconditioners for the Solution of Fully Implicit Multiphase Flow Problems.
SPE Reservoir Simulation Symposium. 2009.

\bibitem[Hu et al., 2011]{FASP}
X. Hu, W. Liu, G. Qin, J. Xu, and Z. Zhang,
Development of a fast auxiliary subspace pre-conditioner for numerical reservoir simulators,
SPE Reservoir Characterisation and Simulation Conference and Exhibition. 2011.

\bibitem[Chen et al., 2006]{Book-Chen}
Z. Chen, G. Huan, and Y. Ma,
Computational methods for multiphase flows in porous media, Vol. 2. Siam, 2006.

\bibitem[Chen et al., 2016]{bos-pc}
H. Liu, K. Wang, and Z. Chen,
A family of constrained pressure residual preconditioners for parallel reservoir
simulations, Numerical Linear Algebra with Applications, Vol. 23(1), 2016, 120-146.

\bibitem[Christie et al., 2001]{SPE10}
M. Christie, and M. Blunt,
Tenth SPE comparative solution project: A comparison of upscaling techniques.
SPE Reservoir Evaluation \& Engineering 4.4 (2001): 308-317.

\bibitem[Wang et al., 2014]{mlp}
B. Wang, S. Wu, Q. Li, X. Li, H. Li, C. Zhang, and J. Xu,
A Multilevel Preconditioner and Its Shared Memory Implementation for New Generation Reservoir Simulator,
SPE-172988-MS,
SPE Large Scale Computing and Big Data Challenges in Reservoir Simulation Conference and Exhibition, 15-17 September, Istanbul, Turkey, 2014.

\bibitem[Wang et al., 2015]{kwang}
K. Wang, L. Zhang, and Z. Chen,
Development of Discontinuous Galerkin Methods and a Parallel Simulator for Reservoir Simulation,
SPE-176168-MS, SPE/IATMI Asia Pacific Oil \& Gas Conference and Exhibition, 20-22 October, Nusa Dua, Bali, Indonesia, 2015.

\bibitem[Peaceman, 1977]{PWM}
D. Peaceman, Interpretation of Well-Block Pressures in Numerical Reservoir Simulation,
SPE-6893, 52nd Annual Fall Technical Conference and Exhibition, Denver, 1977.

\bibitem[Chen et al., 2015]{plat}
H. Liu, K. Wang, Z. Chen, K. Jordan, J. Luo, and H. Deng,
A Parallel Framewrok for Reservoir Simulators
on Distributed-memory Supercomputers, SPE-176045-MS,SPE/IATMI Asia Pacific Oil \& Gas Conference and
Exhibition, Nusa Dua, Indonesia, 20 ¨C 22 October, 2015.

\bibitem[Guan et al., 2015]{pennsim}
W. Guan, C. Qiao, H. Zhang, C. Zhang, M. Zhi, Z. Zhu, Z. Zheng, W. Ye, Y. Zhang,
X. Hu, Z. Li, C. Feng, Y. Xu, and J. Xu,
On Robust and Efficient Parallel Reservoir Simulation on Tianhe-2, SPE-175602-MS,
SPE Reservoir Characterisation and Simulation Conference and Exhibition, 14-16 September, Abu Dhabi, UAE, 2015.

\bibitem[Wheeler, 2002]{mary}
M. Wheeler, Advanced Techniques and Algorithms for Reservoir Simulation, II: The Multiblock Approach in the Integrated Parallel Accurate Reservoir Simulator (IPARS),
The IMA Volumes in Mathematics and its Applications, Springer New York, pp: 9-19, 2002.

\bibitem[Wang et al., 2015]{jcp-bos}
K. Wang, H. Liu, and Z. Chen, A scalable parallel black oil simulator on distributed memory parallel computers,
Journal of Computational Physics, Vol 301, 19-34.

\bibitem[Chen et al., 2009]{inexact-Newton}
T. Chen, N. Gewecke, Z. Li, A. Rubiano, R. Shuttleworth, and B. Yang,
X. Zhong, Fast Computational Methods for Reservoir Flow Models, 2009.

\bibitem[Chen et al., 2016]{lrs2016}
H. Liu, K. Wang, Z. Chen, B. Yang and R. He,
Large-scale Reservoir Simulations on Distributed-memory Parallel Computers,
SpringSim-HPC  2016 April 3-6 Pasadena, CA, USA.

\bibitem[Gander et al., 1994]{golub}
W. Gander, G. Golub, and R. Strebel,
Least Squares Fitting of Circles and Ellipses, BIT Numerical Mathematics,
1994; 34(4): 558-578.

\bibitem[Bank et al., 1989]{ABF}
R. Bank, T. Chan, W. Coughran Jr., and R. Smith,
The Alternate-Block-Factorization procedure for systems of partial differential equations,
BIT Numerical Mathematics, 1989; 29(4): 938-954.

\bibitem[Lacroix et al., 2001]{Quasi-IMPES}
S. Lacroix, Y. Vassilevski, and M. Wheeler,
Decoupling preconditioners in the implicit parallel accurate reservoir simulator (IPARS),
Numerical linear algebra with applications, 2001; 8(8): 537-549.

\bibitem[Cai and Sarkis, 1999]{RAS}
X. Cai and M. Sarkis,
A restricted additive Schwarz preconditioner for general sparse linear systems,
SIAM Journal on Scientific Computing, 1999; 21(2): 792-797.

\bibitem[Falgout et al., 2002]{HYPRE2}
R. D. Falgout, and U.M. Yang, HYPRE: A library of high performance preconditioners,
Lecture Notes in Computer Science, Springer Berlin Heidelberg, 2002. 632-641.

\bibitem[Chen et al., 2016]{chenpoly}
K. Wang, H. Liu, J. Luo, and Z. Chen, Parallel Simulation of Full-Field Polymer Flooding, The 2nd IEEE International Conference on High Performance and Smart Computing, 2016.

\bibitem[Killough, 1995]{spe9}
J.E. Killough, Ninth SPE Comparative Solution Project: A Reexamination of Black-Oil Simulation,
SPE-29110-MS, SPE Reservoir Simulation Symposium, 12-15 February, San Antonio, Texas, 1995.

\end{thebibliography}

\end{document}